%% file: influencemap.tex
\documentclass{vgtc}                          

\ifpdf
  \pdfoutput=1\relax                   
  \usepackage{graphicx}                
  \DeclareGraphicsExtensions{.pdf,.png,.jpg,.jpeg} 
\else
  \usepackage{graphicx}                
  \DeclareGraphicsExtensions{.eps}     
\fi%

\graphicspath{{figures/}{pictures/}{images/}{./}} 

\usepackage{microtype}                 
\PassOptionsToPackage{warn}{textcomp}  
\usepackage{textcomp}                  
\usepackage{mathptmx}                  
\usepackage{times}                     
\usepackage{cite}                      
\usepackage{tabu}                      
\usepackage{booktabs}                  

\usepackage[colorinlistoftodos,prependcaption,textsize=tiny]{todonotes}

\newcommand{\lx}[1]{\todo[linecolor=olive,backgroundcolor=olive!25,bordercolor=olive]{LX:#1}}

\onlineid{0}

\vgtccategory{Research}

\vgtcinsertpkg

\usepackage{amsmath}
\usepackage{multirow}
\usepackage{xcolor}
\usepackage{soul}
\usepackage{xspace}

\usepackage{hyperref}
\hypersetup{
    colorlinks=true,
    linkcolor=blue,
    filecolor=blue,      
    urlcolor=blue,
    citecolor=blue,
}
\newcommand{\eat}[1]{}

\newcommand{\infmap}{{Influence Map}\xspace}
\newcommand{\inflower}{{Influence Flower}\xspace}
\newcommand{\inflowers}{{Influence Flowers}\xspace}

\newcommand{\bA}{{\mathbf A}}
\newcommand{\bC}{{\mathbf C}}
\newcommand{\ba}{{\mathbf a}}
\newcommand{\bs}{{\mathbf s}}
\newcommand{\bS}{{\mathbf S}}

\newcommand{\tbf}[1]{{\vspace{1mm}\noindent {\bf #1}}\xspace}

\newcommand{\smallfigmoveup}{\vspace{-1.0mm}}
\newcommand{\secmoveup}{\vspace{-0.5mm}}
\newcommand{\postfigmoveup}{\vspace{-4.0mm}}
\newcommand{\sectxtmoveup}{\vspace{-1.0mm}}

\preprinttext{To appear in the IEEE Conference on Visual Analytics Science \& Technology (VAST), 2019.}

\title{Influence Flowers of Academic Entities}

\author{
Minjeong Shin, 
Alexander Soen, 
Benjamin T. Readshaw,
Stephen M. Blackburn,
Mitchell Whitelaw,
Lexing Xie\thanks{e-mail: \{{minjeong.shin, alexander.soen, benjamin.readshaw, steve.blackburn, mitchell.whitelaw, lexing.xie}\}@anu.edu.au}\\ %
\scriptsize Australian National University}

\teaser{
  \includegraphics[width=\linewidth]{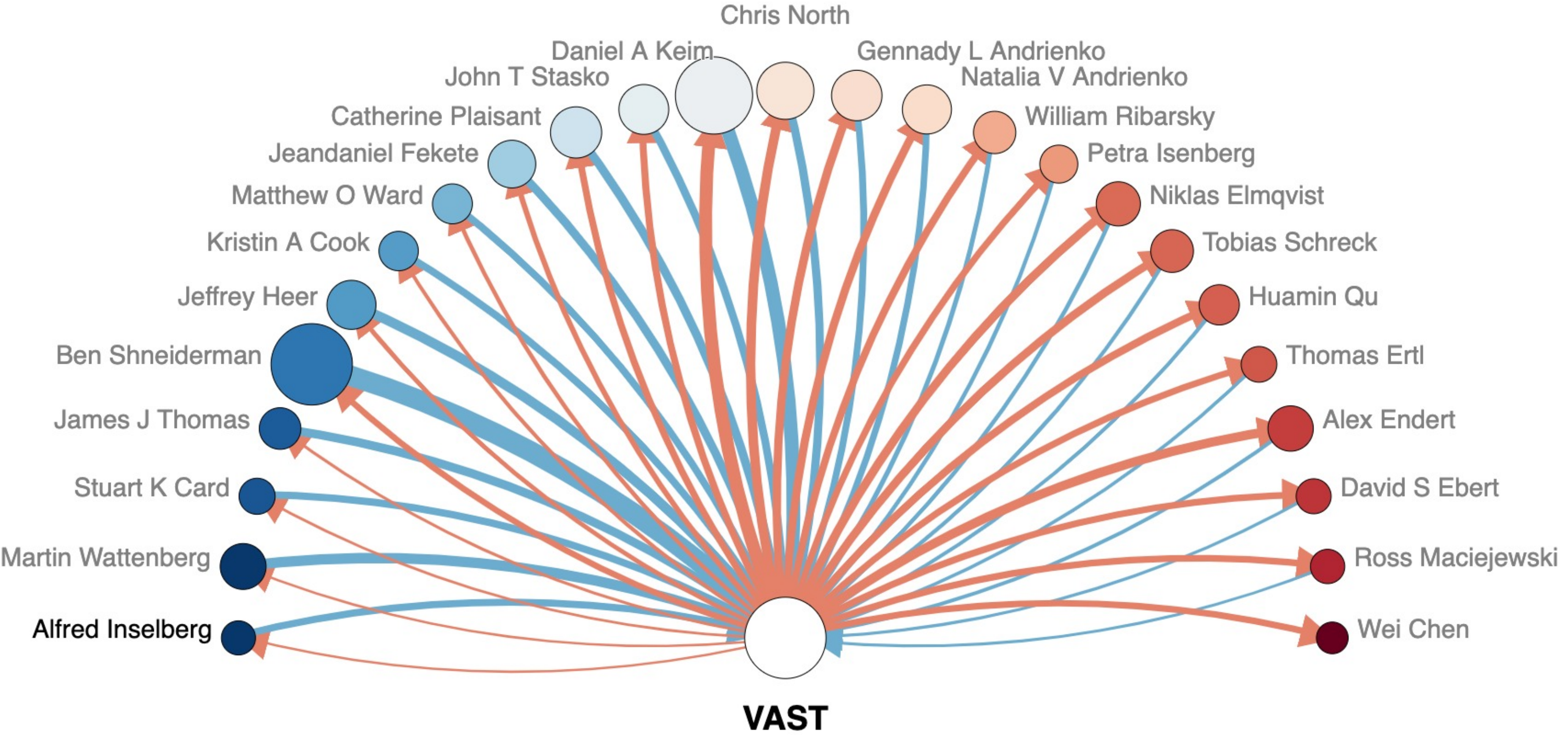}
  \caption{Venue-to-author \inflower of VAST (The IEEE Conference on Visual Analytics Science and Technology). The flower visualises 25 authors who have had the most citation influence to and from VAST. Blue edge width: normalised references to the author's work by VAST papers; red edge width: normalised citations made by the author to VAST papers. Node colour (and sorting) reflect the ratio of references (more blue) to citations (more red); node sizes are scaled by the total amount of references and citations. The 24 authors with names in grey have published in VAST. The only exception is  \href{https://en.wikipedia.org/wiki/Alfred_Inselberg}{Alfred Inselberg}, who is known and being cited for parallel coordinates. }
  \label{fig:main_teaser}\smallfigmoveup
}

\input{sections/abstract.tex}
\usepackage{amsfonts}

\begin{document}



\maketitle

{\tbf{Index Terms:} Human-centered computing -- Visualization --
\{Visualisation application domains -- Visual analytics;
Visualization systems and tools; Empirical studies in visualization.\}}
\input{sections/intro.tex}
\input{sections/related.tex}
\input{sections/dataset.tex}

\input{sections/method.tex}

\input{sections/system.tex}

\input{sections/evaluation.tex}

\input{sections/conclusion.tex}


{\small
\tbf{Acknowledgements} 
This work is supported by ANU College of Engineering and Computer Science, Data to Decisions CRC, and ACM SIGMM. We thank Microsoft and Kuansan Wang for the academic graph data. Computing infrastructure is provided by NECTAR. Nectar is supported by the Australian Government through the National Collaborative Research Infrastructure Strategy (NCRIS). We thank Kathryn McKinley, Bob Williamson, John Lloyd, Thushara Abhayapala, Cheng Soon Ong, Rajeev Gore, Michael Norrish, Joan Leach, Kobi Leins, and Deb Verhoeven for contributing their ideas and perspectives. 
We also thank the anonymous reviewers for constructive suggestions.
}

\newpage

\bibliographystyle{abbrv-doi}

\bibliography{influencemap}

\onecolumn
\newpage
\input{sections/appendix.tex}
\end{document}

%% file: sections/abstract.tex
\abstract{\sectxtmoveup
We present the \inflower, a new visual metaphor for the influence profile of
academic entities, including people, projects, institutions, conferences, and journals.
While many tools {\em quantify influence},
we aim to expose {\em the flow of influence between entities}.
The \inflower is an ego-centric graph, with a query entity placed in the centre.
The petals are styled to reflect the strength of influence 
to and from other entities of the same or different type. 
For example, one can break down the incoming and outgoing influences of a research lab by research topics.
The \inflower uses a recent snapshot of Microsoft Academic Graph,
consisting of 212 million authors, their 176 million publications, and 1.2 billion citations.
An interactive web app, \infmap, is constructed around this central metaphor
for searching and curating visualisations.
We also propose a visual comparison method that highlights
change in influence patterns over time.
We demonstrate through several case studies that the
\inflower supports data-driven inquiries 
about the following: researchers' careers over time;
paper(s) and projects, including those with delayed recognition;
the interdisciplinary profile of a research institution;
and the shifting topical trends in conferences.
We also use this tool on influence data beyond academic citations, by contrasting
the academic and Twitter activities of a researcher.
}

%% file: sections/intro.tex

\section{Introduction}
\label{sec:intro}
\sectxtmoveup

Academic profiles of scientist and organisations are engaging 
for both the scientific community and the general public.
They help us understand individual productivity and reputation,
the collective knowledge-making process, and aid decision-making.
\eat{
\lx{Shall we delete this ORCID example?}
{\color{gray}{For example, ORCID\footnote{\url{https://orcid.org/members}},
the Open Researcher and Contributor ID system,
contain 6+ million profiles and 1000+ member organisations world-wide as of 2019,
-- including academic institutions, industry, funding bodies, publishers, and associations.}}
}
The practice of such understanding
is an active area of research~\cite{fortunato2018science}.
One set of open questions is the
elusive notion of influence,
such as:
How is a researcher or a research project influencing the world,
what ideas did it built on, and who is being influenced?
What is the influence footprint of an organisation,
and how does it change since the founding of a research institute or the inception a conference series?
How does citation influence compare with influence and impact in other means,
such as academic genealogy, mentorship, or social media activities?



The design of visual analytic tools for academic data
has seen much creative energy in recent years.
This paper is motivated by three design considerations.
The first is to focus singularly on influence.
Recent work has studied collaboration~\cite{tang2008aminer,wu2016egoslider,kurosawa2012co},
 popularity~\cite{lee2005paperlens}, and communities inferred from influence~\cite{rosvall2008maps};
or can be multi-focal 
which
include collaborations, topics, and citation relationships~\cite{chen2006citespace,elmqvist2007citewiz}.
We choose to design a visualisation and interaction scheme tailored for understanding influence -- 
rather than splitting the visualisation to accommodate collaboration, popularity, and/or communities additionally.
The second is breadth versus depth.
Online search engines curate all academic data from the web~\cite{sinha2015mag} and
a number of research projects aim to provide a detailed view of a field~\cite{lee2005paperlens,stasko2013citevis,dork2012pivotpaths}. 
Some approaches focus on a single types of entities such as 
papers~\cite{wu2019disrupt,stasko2013citevis}, 
authors~\cite{wu2016egoslider,liu2015egonetcloud},
or keywords~\cite{lee2005paperlens}; while others customise the visualisation for multiple relations~\cite{dork2012pivotpaths} or integrate with other modalities~\cite{latif2019visauthor}.
Here we choose to cover all scientific disciplines and 
aim for a consistent visualisation for all entity types:
from people, to institutions, to publication venues and research topics.
The single focus on bi-directional influence allows for breadth.
The third consideration is expressiveness vs simplicity. 
This work aims to express enough details for users to engage in and understand instances of influence. 
We also choose to focus on a simple relation --
the immediate source and target 
of influence, rather than expanding influence (or collaboration) recursively into a network, 
for which prior work exist~\cite{elmqvist2007citewiz,chen2006citespace}.
The rational for this is that interpreting influence beyond the first hop is difficult for a lay user and 
a network often increases the cognitive complexity for the users. 
The intention is not to create a metric or a set of metrics
as there are many available~\cite{bergstrom2008eigenfactor,bornmann2005does}.

We introduce the \inflower as a new visual metaphor for the bi-directional influence relations between entities (\autoref{sec:measuring}). 
The underlying flower theme is chosen as it
has connotations of intellectual growth and the flourishing of ideas over time.
An \inflower is an ego-centric graph with the ego entity in the centre and alter nodes on a circular arc. 
The flower petals are formed by two curved edges, whose thickness represent the strengths of influence in either direction. The size of alter nodes reflect the total volume of influence and the colour of the alter nodes reflect the ratio of incoming and outgoing influences. 
We quantify influence using citations as the basic unit 
and consider the act of referencing another paper as a signal of incoming influence (to the ego). Similarly, being cited by another paper signals outgoing influence (from the ego).
Normalised citation count is computed using a recent snapshot of Microsoft Academic Graph (\autoref{sec:dataset})
containing publication records since 1800s.

We further construct \infmap\footnote{Code, demo videos, and interactive figures are provided at\\ \url{http://influencemap.ml/vast19}\label{infmap_url}}, an interactive system,
around the \inflower as the main visual element (\autoref{sec:system}). One can search for any entity including authors, institutions, conferences, journals, papers, or aggregate a set of entities to form a project or a group. Enabled by efficient indexing and caching in the back-end, users can sort and filter with the \inflower. One can also compare snapshots of influence over time. An example \inflower is shown in \autoref{fig:main_teaser} and an \infmap snapshot is shown in \autoref{fig:interface_flower}. 

We demonstrate the use of the \inflower and \infmap using a number of case studies (\autoref{sec:cases}), including: visualising a scientist's career corroborated with biographic and interview records; 
picturing the impact of a paper across disciplines (with delayed recognition); 
profiling the intellectual footprint of a research institution;
mapping the topical trend changes in a conference;
and using the \inflower metaphor on non-academic data, by comparing the Twitter and academic influence profiles of a well-known researcher. 

The main contributions of this work are:
\vspace{-\topsep}
\begin{itemize}
    \setlength{\parskip}{0pt}
    \setlength{\itemsep}{1pt plus 1pt}
    \item The \inflower, a new visual metaphor exposing the influence between a wide range of academic entities. 
    \item The \infmap system$^{\ref{infmap_url}}$, available to the public, for curating and interacting with influence flowers in any scientific field. 
    \item Extensive case studies demonstrating data-driven inquires for researchers, research projects, publication venues, organisations, and comparing academic influence to those in social media. 
\end{itemize}
\vspace{-\topsep}

%% file: sections/related.tex
\secmoveup
\section{Related Work}
\label{sec:related}

This work builds on a 
rich literature on academic search engines, bibliographic data visualisation, and studies on the science of science.

\tbf{Academic search engines and libraries}
are modelled after web search engines or library catalogues. A paper and its content are treated as the primary unit for indexing and searching.
Some search services provide entity profiles.
For example, Microsoft Academic~\cite{sinha2015mag}, Scopus,
and AMiner\cite{tang2008aminer} contain profiles of authors, affiliations, and/or research fields; and authors can enable and curate their own page on Google Scholar.
Such profiles typically focus on three types of data: productivity -- such as papers published over time, broken down by venue; collaboration -- such as coauthor list or networks; and influence -- via proxies such as the total number of citations, often broken down over time.
A few systems offer in-depth analysis by joining citations with other sources of data.
Semantic Scholar builds influence scores
by distinguishing important vs unimportant citations~\cite{valenzuela2015identifying}.
Altmetric\cite{konkiel2015altmetrics} combines citation metrics with other online sources to derive a single influence score for each paper.

To the best of our knowledge, most existing systems focus on quantifying influence rather than exposing the flow of influence between entities of various types.
Of particular relevance is Semantic Scholar, where the author influence pane shows the top 5 influencers (\autoref{fig:semantic_scholar} in the appendix). We feel that top five entities is a nice teaser but not an instrument for understanding influence. Furthermore, the presentation is a simple flow diagram, lacking a salient visual metaphor.


\tbf{Visualisation of bibliographic data} has long fascinated visualisation
researchers as they are rich, multi-relational, and in a domain that researchers can
readily relate to.

There are many engaging visualisations made by the InfoVis community using
the publications of InfoVis and HCI,
such as the PaperVis~\cite{chou2011papervis}, PaperLens\cite{lee2005paperlens}, PivotPaths\cite{dork2012pivotpaths}, and CiteVis\cite{stasko2013citevis} systems.
All four 
systems contain an overview of papers and topics. Further, they either support users explorßing topical trends\cite{lee2005paperlens}, allow drilling down on paper-author-topic relations\cite{dork2012pivotpaths}, or present the citation relationships between papers~\cite{chou2011papervis, stasko2013citevis}.
These systems are paper-centric and require a predefined community or domain
to start the overview and exploration. It can be hard to use such a methodology
on a large scientific domain (e.g. astrophysics) which has too many papers to fit
on a screen and papers scattered around a large set of journals
(from {\em Nature}, {\em Science} to {\em Icarus} and 
{\em Astronomy \& Astrophysics}).

The network of coauthors and ego-centric views of one's collaboration
have also been a prevalent theme for visualisations.
Such systems are often designed with different emphases and goals,
such as quantifying the time and strength of mutual influence~\cite{reitz2010framework},
predicting researchers’ future activities~\cite{kurosawa2012co},
and depicting coauthorship with a subway map metaphor\cite{zhao2016egolines}.
One particular foci of multiple systems has been to track collaborations over time,
such as the design of egoSlider~\cite{wu2016egoslider},  EgoNetCloud~\cite{liu2015egonetcloud},
and 1.5D ego network visualisation~\cite{shi201515d}.
For these systems, the notable limitations are that the ego networks are focusing on authors
as the key entity. The corresponding visual metaphors
do not readily generalise to other academic entities.

Besides the two topical clusters above,
a diverse set of visual paradigms have been explored on academic data.
Fung et al.~\cite{fung2016design} present a design study
for the bibliographic record of a person,
multiple attribute-mapping schemes which are applied to
three visualisations schemes: networks, trees, and matrices.
Wu et al.~\cite{wu2019disrupt} use the tree metaphor to
represent a paper's heritage and influence over time.
Latif and Beck~\cite{latif2019visauthor} present a text-centric summary
of one's scientific career, enriched by sparklines (inline mini figures),
and side panels on collaborator networks and coauthored work.
A number of systems consist of multiple connected information-dense visualisations.
PaperLens has different views on topics and trends in research fields~\cite{lee2005paperlens}.
CiteWiz~\cite{elmqvist2007citewiz} portrays a network of authors and topics,
along with glyphs for author influence over time.
The CiteSpace system~\cite{chen2004searching,chen2006citespace} integrates a
rich set of visualisations with network analysis; driven
by the goal of finding research fronts and emerging trends~\cite{chen2006citespace},
and finding turning points using a co-citation network \cite{chen2004searching}.
Two notable recent visualisation systems, egoSlider~\cite{wu2016egoslider} and ImpactVis~\cite{wang2018visualizing}, propose rich visualisations of collaboration and influence over time. Both systems treat time as an essential dimension, either in the stream of collaborations~\cite{wu2016egoslider}, or as one dimension in the main influence matrix~\cite{wang2018visualizing}. \inflower choose not to express time in the main visual metaphor, but offers it as a data filtering option instead (\autoref{sec:system}). 

\tbf{The science of science}~\cite{fortunato2018science}
is an active research area that uses large data sets to study the mechanisms underlying
the production of new knowledge.
In terms of identifying highly influential work,
Sinatra et al.\cite{sinatra2016quantifying} observed that the highest impact
work in scientist's career
are
randomly distributed,
and Wu et al.\cite{wu2019disrupt} found that the age of related work and team size
are correlated with producing disruptive work.
Van Rann~\cite{van2004sleeping} identified papers with delayed recognition
as a common phenomena. Ke et al.\cite{ke2015defining} proposed methods to
systematically identify them from citation time series.
Hoonlor et al.~\cite{hoonlor2013trends} mapped trends in computer science and quantified
the fraction of keywords in papers that trend before or after they do so in grant applications.
Another line of inquiry examines scientists' demographics and scientific achievements.
Sugimoto et al.\cite{sugimoto2017scientists} found a positive correlation
between scholars' scientific impact and their mobility;
Lariviere et al.\cite{lariviere2013bibliometrics} profiled gender disparities in science; and King et al.\cite{king2017men} observed different citation practices between men and women.


In summary, this work contributes one central  metaphor
for visualising the ego network of diverse types of entities.
Our focus is on influence rather than collaboration.
The goal is to broadly cover all academic fields.
We aim to enable data-driven formulations and answers to questions in science-of-science,
using the ego-centric \inflower metaphor and its supporting interactions.

%% file: sections/dataset.tex
\secmoveup
\section{Dataset}
\label{sec:dataset}
\sectxtmoveup

We use the Microsoft Academic Graph (MAG) dataset to compute influence statistics. MAG is the data source behind Microsoft Academic\footnote{https://academic.microsoft.com}
and is a large and openly available academic dataset covering all research fields. MAG is comprised of six types of academic entities and their relations~\cite{shen2018magfos}: paper, author, institution, venue (journal and conference series), event (conference instance), and topic. 
Entity types of paper, author, institution, journal, and conference are discovered from structured (e.g.\ publisher and knowledge base) and unstructured (e.g.\ web pages indexed by a search engine) sources. MAG provides entity resolution such as merging records from different sources, de-duplication, and disambiguation.

Microsoft Academic uses a large vocabulary of research topics\cite{sinha2015mag}, seeded by Wikipedia entries, to classify each of its papers. These topics are mapped to papers using a machine learning algorithm on paper information, such as title, abstract, and publishing venue.
MAG includes 230K
research topics organised into a six-level hierarchical structure. Level 0 includes nineteen top-level disciplines, such as physics and medicine. Topics are organised into a directed acyclic graph.  An example chain of parent-child relations might be: 
{\em {computer science}, {algorithms}, {computational complexity theory}, {NP-complete}.}
Level 0 and level 1 concepts are manually reviewed. In this study, we use 294 level 1 concepts to measure topical influence. 

The current \infmap system is based on a MAG data snapshot taken on 2018-06-29. It includes scientific papers published from 1800 to 2018. The dataset has 176 million articles, 212 million authors, and 52 thousand journals and conference series. It also includes 468 million paper-author relationships, 1.2 billion paper-to-paper references, and 949 million paper-to-topic mappings. About 70K papers are added to MAG daily\cite{shen2018magfos}, we plan to update the \infmap system with newer MAG data periodically.



\tbf{A discussion on data coverage and quality.}
Results of large-scale information extraction, entity resolution, and classification from the web is not necessarily perfect (e.g., Appendix \autoref{fig:brian_medicine}). 
We briefly review recent work that validate the coverage and accuracy of MAG, and present a few of our own observations. In addition, our interactive system allows users to correct entity resolution errors by merging MAG entity ids (\autoref{ssec:searching}).

Microsoft Academic has been compared to other popular bibliographical sources. 
On a sample of 145 academics across five disciplines~\cite{harzing2017microsoft2}, MAG was found to have better coverage than Web of Science and Scopus, but less coverage in book chapters and miscellaneous publications (white papers, newletter articles) than Google Scholar. Another work builds on a verified 
publication list from an entire university~\cite{hug2017coverage}. MAG was found to share the same bias as Scopus and Web of Science to 
cover less humanities, non-English, and open-access publications.

Microsoft Academic reports two versions of citations counts for each publication -- a count of algorithmically verified citations, 
but also an estimated citation count. There are no ground truth citation counts to compare with, however it has been observed that the verified citations tend to be lower than what Google Scholar reports~\cite{harzing2017microsoft2}. 
Sometimes inflated and deflated citation counts are found in MAG due to document merging~\cite{Hug2017citation}. 
Furthermore, the practice of preprinting has made
accurately dating the publication more difficult -- 
although Google Scholar suffers from similar issues. 

We observe two phenomena in the computer science research community that may affect citation statistics.
First, articles appearing within a given venue (especially conferences) may be subject to entirely different editorial processes. For example, a single conference proceedings may include full-length rigorously reviewed articles, poster abstracts, demos, and panel sessions without metadata identifying these distinctions. \autoref{fig:chi_papercount} in the appendix illustrates this using ACM SIGCHI as an example. This may dilute 
the apparent impact of a prestigious peer reviewed venue that also contain large numbers of short, poorly cited ancillary articles such as demos and posters.
Second, papers are sometimes published as a short conference abstract and then later as a full paper, or sometimes simultaneously as conference and journal papers (e.g. VAST and TVCG). 
This leads to ambiguity in resolving the publication venue of a paper, and will affect aggregated statistics.
One strategy for addressing this issue, used by a recent ranking system~\cite{berger2019goto}, is to query MAG with curated paper titles (e.g. from DBLP). They found that 97\% of papers from 308 computer science venues have corresponding entries in MAG.


%% file: sections/method.tex


\secmoveup
\section{Measuring intellectual influence}
\label{sec:measuring}

We describe the composition and visual design of an \inflower, the approach for computing influence scores, and a proposed method to visually compare two flowers.

\secmoveup
\subsection{The \inflower}
\label{ssec:flower}

We use citations as a proxy for intellectual influence and adopt a simple method to quantify it. If paper 1 cites paper 2, then paper 2 is considered to have influenced paper 1. Note the direction of influence is the opposite of the citation. Other data sources of influence exist, such as Altmetric\cite{konkiel2015altmetrics}, academic genealogy\cite{lienard2018intellectual}, or unstructured interviews (e.g. \cite{shafi_interviews}). 
We leave incorporating other data sources of influence as future work.

We design the \inflower, a new visual metaphor for presenting aggregated influence around a given academic entity. Academic entities could be a paper, a project, an author, an institution, a conference, a journal, or a topic. An academic entity is represented as a collection of papers in MAG. If it is an author, it is the collection of papers that the author has authored. If it is a conference series, it is the collection of published proceedings of the conference.

The \inflower is an egocentric graph, with one node in the centre (the ego), and other related nodes on the outside (the alters).
We define a node, either the ego or alter, as an academic entity.  Edges in the flower indicate the influence relation between the ego and the alters. The direction of an edge denotes the direction of the influence.
The curved edges between the ego and alters form petals of the flower that blossoms when animated. 
We omit the edges between alters to preserve the visual layout of a flower. The flower is designed to support visual inquiries (\autoref{sec:cases}) about the relationships between the ego and its alters, rather than the network around the ego.

Other metaphors have been introduced to visualise academic entities.
The closest to the current one is botanical trees~\cite{fung2016design, wu2019disrupt},  where tree height represents the time, and leaves and roots represent citations and references.
Streams is another metaphor for tracking changes in collaboration over time \cite{liu2015egonetcloud, wu2016egoslider, zhao2016egolines, shi201515d}. 
The key focus of the \inflower is on the aggregated {\em strength} of influence between entities,
whereas trees and streams use time as a primary dimension. 
With the \inflower, temporal changes can be captured by snapshots of the flower at different points in time, as we describe in \autoref{ssec:compare}. 


\secmoveup
\subsection{Computing influence scores}
\label{ssec:infscore}\sectxtmoveup

To generate nodes and edges of an \inflower, we aggregate the pairwise influence of entities associated with the papers of the ego.
We regard as a unit of influence a citation made by a single paper and received by another, since this is the smallest unit reflected in academic data.

We denote $E$ as the set of entities and $P$ as the set of papers in the dataset. We consider six entity types: 
\text{author}, \text{venue}, \text{institution}, \text{topic}, \text{paper}, and \text{project}. 
The association matrix $\bA  \in \{0,1\}^{|E| \times |P|}$ indicates the relations between entities and papers. 
The corresponding association matrices of author, venue, institution, and topic types are $\bA^{(a)}$, $\bA^{(v)}$, $\bA^{(i)}$, $\bA^{(t)}$. That is to say, element $A_{ij}^{(a)} = 1 $ if and only if paper $j$ is authored by person $i$, 0 otherwise. Note that row vector $\ba_{i\cdot}^{(a)}$ has value 1 for all papers authored by person $i$ and column vector $\ba_{\cdot j}^{(t)}$ has value 1 for all topics that paper $j$ is relevant to (see \autoref{sec:dataset}). Projects and ad-hoc paper collections are encoded as a vector of indicators, i.e. a one-hot vector for a paper and a multi-hot vector
for a project. For example, $\ba^{(p)} = [1, 0, \ldots, 0, 1]$ refers to a paper collection containing two papers, at the very beginning and end of the paper set, respectively.

Citation matrix $\bC \in \{0,1\}^{|P| \times |P|}$ represents the citation relation between papers.
$C_{ij} = 1$, if $p_i$ influences $p_j$ ($p_j$ cites $p_i$), otherwise $0$. 
Note that we do not compute the entire citation matrix $\bC$. The index structure in \autoref{ssec:indexing} is used to obtain the relations between the ego entity and its references and citations, which are represented by a column and a row of $\bC$.

The total influence can be expressed as multiplication of these indicator matrices. To obtain influence scores between all entities of type $k$ and type $l$, where $k, \, l \in {\{a, \, v, \, i, \, t\}}$, we only need to compute: 
\begin{equation}
    \bS^{(k,l)} = \bA^{(k)} \bC \bA^{(l)T} 
    \label{eq:infscore}
\end{equation}
Hence, the influence score between author $i$ and topic $j$ is calculated as $S_{ij}^{(a,t)} = \ba^{(a)}_i \bC \ \ba^{(t)T}_j .$ 

For example, the influence scores from author $i$ to all authors is represented as a vector 
$\bs_{i\cdot}^{(a,a)} = \left[S_{i0}^{(a,a)}, S_{i1}^{(a,a)}, \ldots \right]$; 
the influence vector from all authors to venues $j$ is 
$\bs_{\cdot j}^{(a, v)} = \left[S_{0j}^{(a,v)}, S_{1j}^{(a,v)}, \ldots\right]^T$.

A sorted and truncated set of scores in $\bs_{i\cdot}^{(a,a)}$ can be rendered as the red (outgoing) influence edges in an author-to-author \inflower (e.g. \autoref{fig:shafi}) and $\bs_{\cdot j}^{(a,v)}$ would be blue edges in a venue-to-author \inflower (\autoref{fig:main_teaser}). Faster computation of influence scores are implemented using indexing techniques in \autoref{ssec:indexing}.







A normalisation scheme for influence scores is crucial for its use and interpretation, especially since modern academic papers can have between one and thousands of authors, and cite between a handful to a few hundred other works. 
We normalise the influence score so that one citation has one unit of influence.
\autoref{sec:normalisation} discusses the normalisation method, several alternatives investigated and a pilot validation.

\secmoveup
\subsection{Visualising an \inflower}
\label{ssec:infvis}
\sectxtmoveup

\tbf{An \inflower}
is laid out on a circle
with the ego in the centre and alters evenly distributed along the arc. 
We choose a circular design to make flower petals of the same length. Alternatives, such as placing alters in a horizontal line break the flower metaphor and could mislead due to the different edge lengths.
We decrease the angular span if the flower has less than 10 petals
and linearly increase it if there are more than 25 petals.
The maximum span is 270 degrees, for a maximum number of 50 petals.
The alters with the highest maximum of incoming and outgoing influence are selected. We sort the alters by the maximum of two influence instead of the sum to avoid cases when one colour overwhelms another. 
The default number of petals is 25. 
The selected petals are sorted by the influence ratio (blue to red colour) by default.
The number and the ordering can be adjusted interactively (\autoref{sec:system}).
The scale of the flower is adjusted according to the display size and the number of petals. 

\tbf{Edge} appearance is determined by the influence score.
The colour and arrowhead of an edge indicate its direction. We assign two contrasting colours, red to denote the influence that the ego has towards the alters and blue to represent the influence that the alters have towards the ego.
The weight of the edge $w_{ij}$ denotes the strength of influence from entity $e_i$ to $e_j$. $w_{ij}$ is proportional to the normalised influence score $\bar{S}_{ij}$ (\autoref{sec:normalisation}) and log-normalised for visualisation.

\tbf{Node} colour and size are determined by the pair of edges connected to the node.
The colour of a node signifies the difference in strength between incoming and outgoing influence. We define influence ratio as the difference between the incoming and outgoing influence, normalised by their sum.
A blue (\#053061) to red (\#67001f) interpolator, \href{https://github.com/d3/d3-scale-chromatic#interpolateRdBu}{d3-scale-chromatic}, is used to determine node colour according to the influence ratio.
The ego node is white. 
The size of a node is proportional to the sum of influence.
We scale the maximum node size according to the display size.
The ego and the biggest alter have the maximum size, while the other nodes are scaled accordingly.

\begin{figure}[t]
  \centering
  \includegraphics[width=0.45\textwidth]{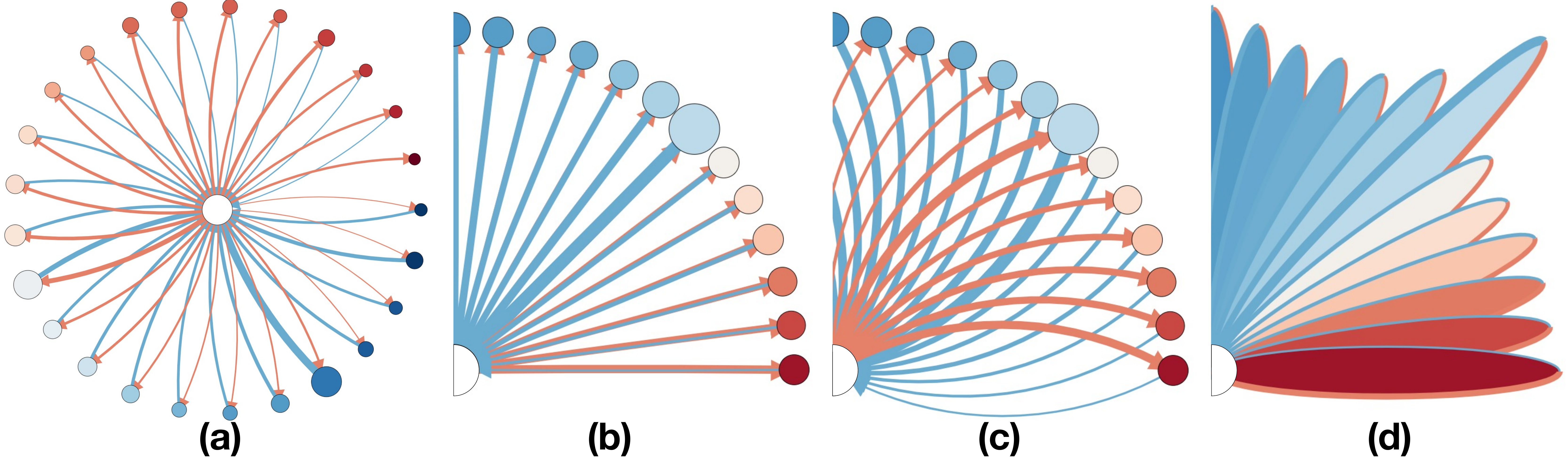}
  \caption{Alternatives designs for the influence flower and petals. (a) A full circle flower design. (b) Flower petals with arrows in the both ends. (c) A wide band of flower petals. (d) Coloured flower petals.
  }
  \label{fig:design_alts}\postfigmoveup
\end{figure}

\tbf{Alternative designs} of the \inflower
are shown in \autoref{fig:design_alts}. 
A full circle flower (\autoref{fig:design_alts} (a)) allows a larger number of alters and edges are less overlapped compare to a semicircle flower. However, the semicircle (or incomplete arc) reinforces the flower metaphor as it has a sense of gravity where the ego at the bottom can be seen as a flower stem. Moreover, the left to right arrangement in the semicircle makes it easy to understand the petal order.
We considered different shapes of flower petals (\autoref{fig:design_alts} (b), (c) and (d)). Straight edges (b) are not preferred because they do not effectively depict a flower and it is hard to discern the size of incoming and outgoing influence. The earlier version of the influence flower had a wider band of petals (c), but were changed to a narrower band for better readability. We also considered using the colour of a filled petal (d) to indicate influence ratio and radius to indicate influence strength. It was not chosen since
concurrently varying edge, fill, and length makes interpretations harder and different petal lengths adversely impacts the label layout.

In the current flower layout, displaying more than 50 petals is difficult. 
One could potentially scale up the display using other flower shapes,
such as layered (e.g. roses) or with a geometric space partition (e.g. pompom dahlias).
Another direction for extending the metaphor can be multiple flowers in a bouquet (e.g. corresponding to a research group) or a garden (e.g. individuals in an institution), where each flower can be examined in detail on demand.

\tbf{Sorting and filtering} provide various perspectives from which to analyse \inflower.
Four different sorting options, influence ratio (node colour), influenced by (blue edges), influencing (red edges), and the total influence (node size) are provided for interaction (\autoref{ssec:interacting}).
Each method changes the ordering of alters in a flower, but it does not change the selection of the top alters.

We consider two filtering options: co-contributors and self-citations. 
The \inflower is able to capture less obvious influence outside of one's co-author networks with the filtering. 
We define two entities to be co-contributors if the entities have contributed to the same paper. For the venue type entity, co-contribution indicates if the ego has published a paper to the venue. For the topic type entity, it means that the ego has written a paper of the topic. Co-contributors of the ego are indicated by nodes with greyed out names.
We define a self-citation between a paper and a cited paper as a relation dependent on the ego. A paper citation is a self-citation if both papers have the ego as an author (a venue, an institution, or a topic).
Citations from co-authors are included
by default, because the focus is the ego node and influencing co-authors (who likely have a separate intellectual profile and affiliation) are an integral part of one's influence profile.

\autoref{fig:main_teaser} is an example venue-to-author \inflower of the IEEE Conference on Visual Analytics Science and Technology (VAST). The ego of the flower, VAST, consists of papers published in this conference since its inception. The alter nodes of the flower are 25 authors with the highest combined incoming and outgoing influence, ordered by node colour. 
Ben Shneiderman is the most influential author to VAST (rank 1 by blue edge score and combined score).
Daniel A. Keim is the author most influenced by VAST (rank 1 by red edge score) and also the second most influential to VAST (rank 2 by blue edge).
Most authors publish in VAST (name in grey font). The exception is Alfred Inselberg, who is cited in VAST for his work on parallel coordinates, and has been influenced by VAST papers. 


\begin{figure}[t]
  \centering
  \includegraphics[width=0.48\textwidth]{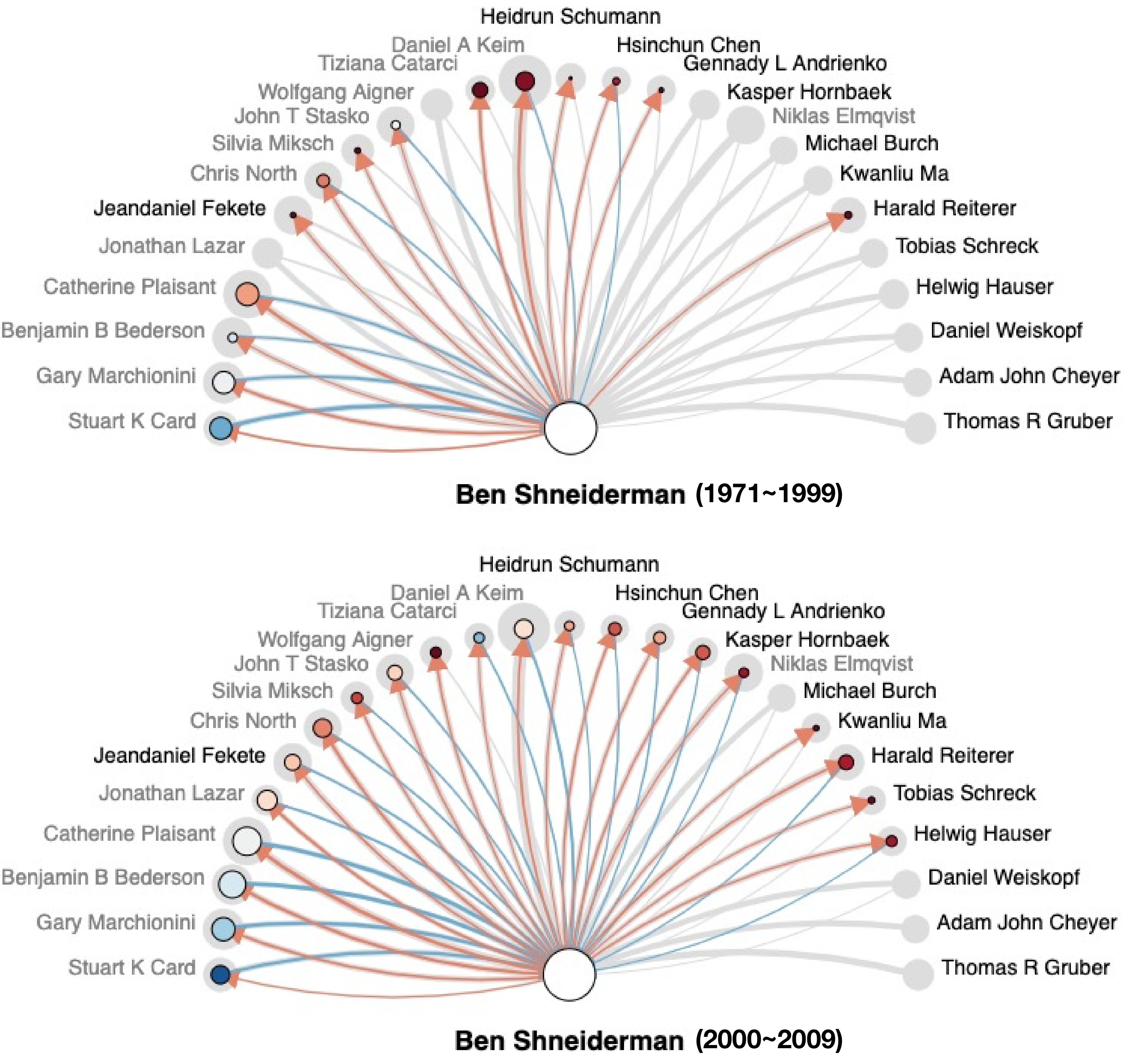}
  \caption{Comparing author-to-author \inflower of Ben Shneiderman using publications and citations in three time periods. Top: 1971 to 1999. Bottom: 2000 to 2009. Grey background: anchor flower from 1971 to 2018. See \autoref{ssec:compare} for discussions.}
  \label{fig:ben_shneiderman}\postfigmoveup
\end{figure}

\subsection{Comparing \inflowers}
\label{ssec:compare}
\sectxtmoveup

To compare multiple \inflowers generated at different time periods, we created the concept of an anchor flower. The anchor flower is the reference against which an \inflower (the contrast flower) is compared. The anchor and contrast flowers share the same ego and alters in the same type.
The time period of the anchor is always the superset of all contrast flowers 
-- here the entire time range of the ego is used.
The anchor is greyed out in the background, and a contrast flower is drawn in colour on top of the anchor flower. 
The anchor flower determines the overall node ordering, background size and position. 
The node size and edge weight in a contrast flower is calculated by the relative influence score with the corresponding node and edge in the anchor. 
The colour of the node is decided by the contrast flower while keeping node order and position fixed by the anchor flower.
Note that nodes in the contrast flower are a subset of the nodes in the anchor. 
We stack two flowers to allow direct size comparison, unlike other studies \cite{zhao2016egolines, wu2016egoslider} arranging ego-graphs in a sequence to represent temporal dynamics.
Comparisons between two different entities, two different types, or time periods which are not subsets of one another is left as future work.

\autoref{fig:ben_shneiderman} compares two snapshots of the author-to-author \inflowers of Ben Shneiderman, the largest alter node in \autoref{fig:main_teaser}. 
The anchor flower is created using his career publications and citations 1971-2018, greyed out as a background.
The top flower is created using publications and citations from 1971 to 1999, the bottom flower 
using those from 2000 to 2009. 
Catherine Plaisant is the largest alter in both flowers. 
For Shneiderman, Plaisant bore more outgoing influence before 2000 (red), 
the influence almost equalised in the 2000s (white). 
One may also notice Niklas Elmqvist, whose mutual influence with Shneiderman was non-existent before 1999, began in the 2000s and mostly happened after 2009. This is corroborated by career information obtained from Elmqvist's homepage -- that his first paper was published after 1999, he started his faculty career between 2000 and 2009 (2008), and he joined the University of Maryland, where Shneiderman works, in 2014.

%% file: sections/system.tex
\secmoveup
\section{The \infmap System}
\label{sec:system}
\sectxtmoveup

\begin{figure}[t]
    \centering
    \includegraphics[width=0.46\textwidth]{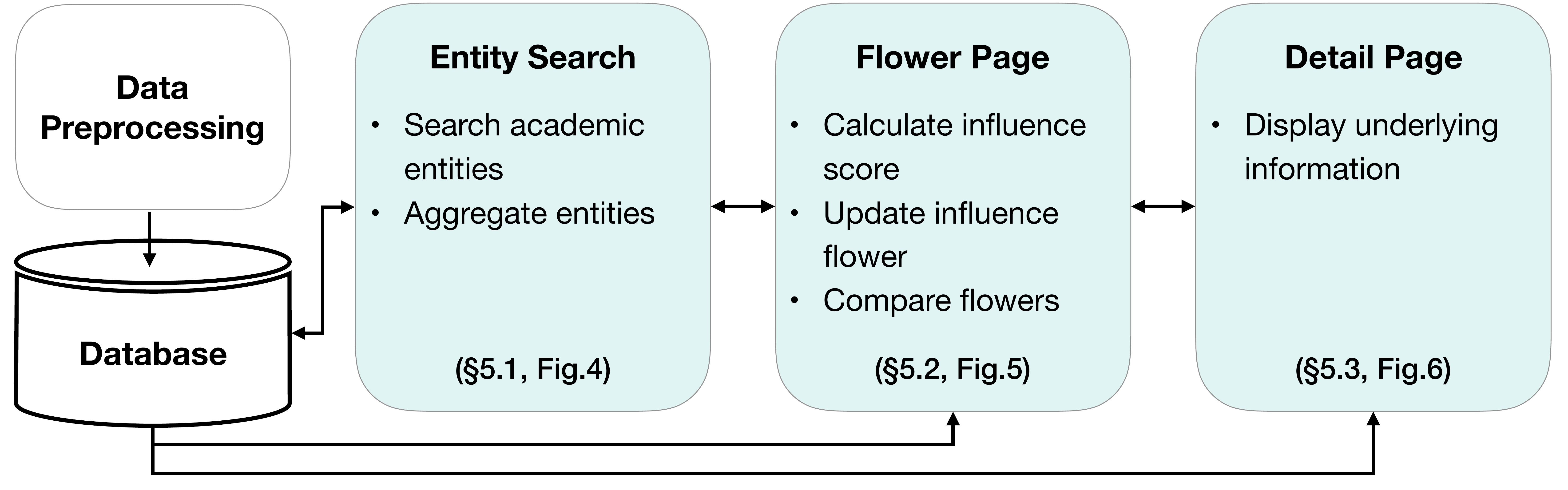}
    \caption{Overview of the \infmap system summarising the main interactions in each component. Snapshots of the entity search page, the main flower page, and the detail page are in \autoref{fig:search_page}, \autoref{fig:interface_flower}, and \autoref{fig:detail_page}, respectively.}
    \label{fig:pipeline}\postfigmoveup
\end{figure}

The \infmap system is built around the \inflower as the central visual component and 
it contains three additional interactive components: searching and curating a flower, the flower visualisation module, and the details page. \autoref{fig:pipeline} presents the flow of the system.
Searching and curating entities is done by querying via Elasticsearch. 
All the scoring and data manipulation is done in Python 3 with the pandas library.
The web user interface is implemented with HTML and JavaScript and styled
using Bootstrap. We use D3.js\cite{bostock2011d3} for the visualisation of charts and flowers.
In \autoref{ssec:searching}--\autoref{ssec:edge}, we explain each step of the workflow of the \infmap system. \autoref{ssec:indexing} describes indexing and caching strategies to scale the \inflower to tens of thousands of citing papers. 

Common entry points to a visual analytics system include browsing and search. Here we choose search as the main entry point. On one hand, current-day users are accustomed to web search and scholar search; on the other, making an overview for hundreds of millions of entities is a design challenge on its own and lies beyond the scope of the current work.
As an alternative entry point, we curate the \infmap gallery with a set of authors, venues, and projects to help users' initial engagement with the system without the two-step search process. A screenshot of the gallery is shown in \autoref{sec:gallery}.

\secmoveup
\subsection{Searching for an academic entity}
\label{ssec:searching}
\sectxtmoveup

\begin{figure}[t]
    \centering
    \includegraphics[width=0.46\textwidth]{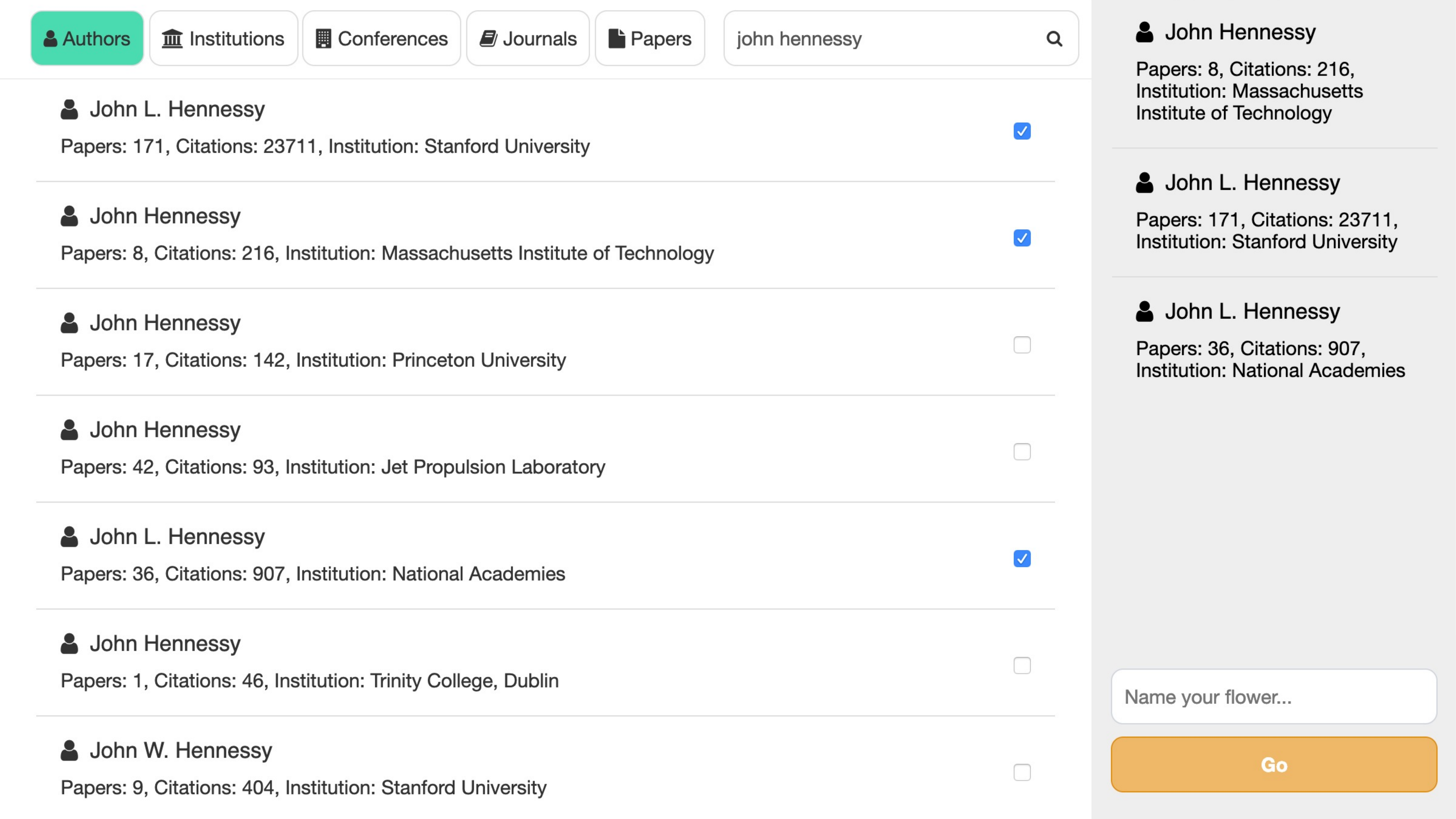}
    \caption{Search for the author name, John L. Hennessy. The number of papers, citations, and affiliation information help find the correct entities of interest.}
    \label{fig:search_page}\postfigmoveup
\end{figure}

Creating an \inflower starts with searching for names of academic entities, which may be one or more of authors, institutions, conferences, journals, and papers. 
Elasticsearch supports full-text query and query results are sorted by their relevance score.
The full-text query is especially useful for searching close variants of author names, which may be abbreviated or missing middle initials.
Furthermore, we modify the relevance score to favour entities with higher citation counts.  \autoref{ssec:relevance} discusses the modified scoring function.


\autoref{fig:search_page} shows the user interface of the search page.
The search page consists of three components: a search bar (top), a result table (left), and a curation list (right).
Using the search bar, users first select the type(s) of an entity to search by name. Then the search result appears in the result table. The number of papers and citation information is provided for each entity. We additionally provide affiliation information for searching authors and author information for searching papers to reduce ambiguity. 
The checkbox on the right-hand side of a row in the results is used to add the entity to the curation list. Selected entities can be removed by un-ticking the checkbox or clicking on the selection on the curation list. The system allows aggregation of multiple entities of different types, which can be used for selecting multiple entity ids with the same name: creating a project flower by selecting multiple papers or authors, or creating a flower with the sum of related conferences and journals. Users are able to rename the flower of aggregated entities.
Finally, the 'go' button on the bottom leads to the \inflower page for the curated entity list.


\secmoveup
\subsection{Interacting with the \inflower}
\label{ssec:interacting}
\sectxtmoveup

\begin{figure*}[t]
    \centering
    \includegraphics[width=\textwidth]{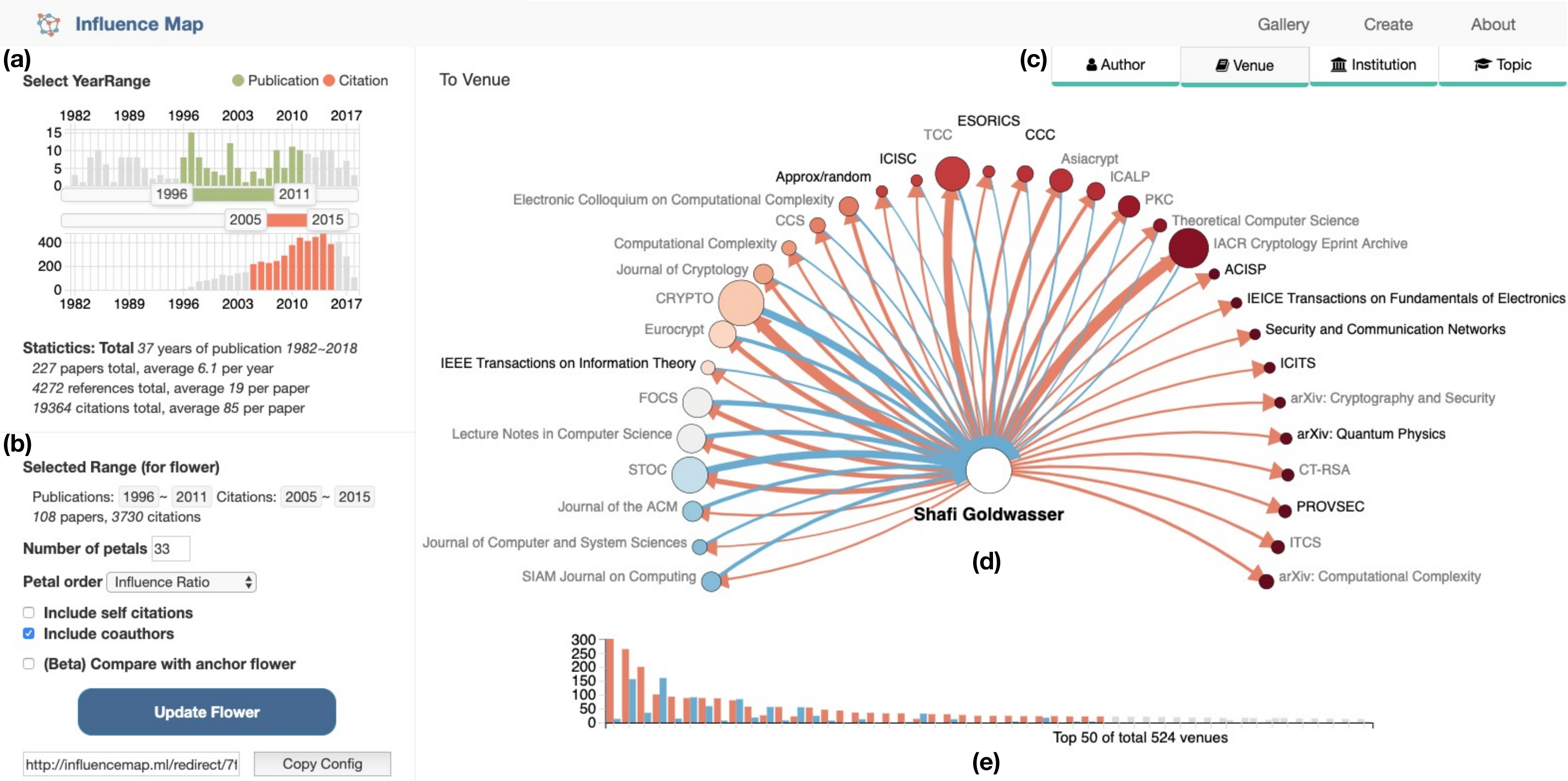}
    \caption{Snapshot of the \infmap system containing an author-to-venue \inflower. The ego entity is Shafi Goldwasser, 2012 Turing Awardee for foundational work on modern cryptography. The alter nodes are publication venues, with conferences shown as acronyms and journals as full names.
    The system consists of (a) year range filter and statistical summary, (b) fine-grained control, (c) influence type tabs, (d) \inflower, and (e) influence overview bars. See \autoref{sec:system} for a description of system components and \autoref{ssec:career} for a discussion on Goldwasser's influence profile.
    }
    \label{fig:interface_flower}\postfigmoveup
\end{figure*}

\autoref{fig:interface_flower} shows the user interface of the \inflower page. The main interface consists of five components: a year range filter and statistics panel, a fine-grained control panel, influence type tabs, the \inflower, and influence overview bars.
The \inflower of the searched entity (\autoref{fig:interface_flower} (d)) is located in the middle. The influence type tabs (\autoref{fig:interface_flower} (c)) switch from four different types of flowers: author, venue, institution, and topic.

The influence overview bars (\autoref{fig:interface_flower} (e)) provide a complementary way to understand influence. It presents the number of references and citations of the top 50 entities, sorted by the maximum of incoming and outgoing influence. We choose alters based on the max rather than total influence, because we would like to present a somewhat balanced view of incoming and outgoing influence -- total influence is more likely to be dominated by one colour.
The overview bars show
the coverage of the \inflower. The entities appeared in the \inflower are drawn in colour, otherwise in grey. The total number of entities is also presented.
Each red and blue bar maps to a corresponding red and blue edge in the \inflower. Mouse-overing a node in the \inflower highlights the connecting arrows and the corresponding location in the influence bar chart, and vice versa.
While the \inflower focuses on visualising the relative differences between nodes and edges, 
the overview bar chart helps users understand absolute values underlying the flower metaphor. This also  captures the maximum number of references the ego made to the alters (and citations the ego received from the alters).

The default \inflower is created to cover the entire academic time span of the ego. We define the entire academic time span as the period from the first publication record to the last citation record in the database. The number of petals by default is 25, and petals are sorted by influence ratio (blue to red colour) by default.
Users are able to change the time range and properties of the \inflower via two filters on the left.

The year range filter (\autoref{fig:interface_flower} (a)) enables filtering by time
using two bar charts: a publication chart (top, green) and a citation chart (bottom, orange) that share the same x-axis range.
Each bar in the publication chart indicates the number of papers that the ego published in the given year. Similarly, bars in the citation chart indicate the number of citation that the ego received. 
The publication range slider (green) under the publication chart allows users to select a specific publication time span of interest. 
The citation chart will be recalculated using the papers in the selected publication year range. 
Users can further filter the received citations using the citation range slider (orange) to specify the time span in which papers are cited.
The statistics panel below the year range filter 
presents an overall summary, showing the average and total number of papers, of the references and citations generated in the entire academic career of the ego. 

The fine-grain control panel (\autoref{fig:interface_flower} (b)) presents the properties required to create or update an \inflower. First, it shows the number of papers and citations selected from the year range filter. 
The system then provides five options to alter the flower: the number and sorting order of petals, options to toggle inclusion of self-citations and co-contributors, and an option to compare the flower with the anchor flower.
Pressing the `update flower' button will create the new flower. The system initially shows the flower with the entire year range and default options. 
Note that the `copy config' 
button on the bottom left of the page provides a link with the current flower configuration for sharing and later access. 

\secmoveup
\subsection{Details behind an edge}
\label{ssec:edge}
\sectxtmoveup

\begin{figure}[t]
    \centering
    \includegraphics[width=0.48\textwidth]{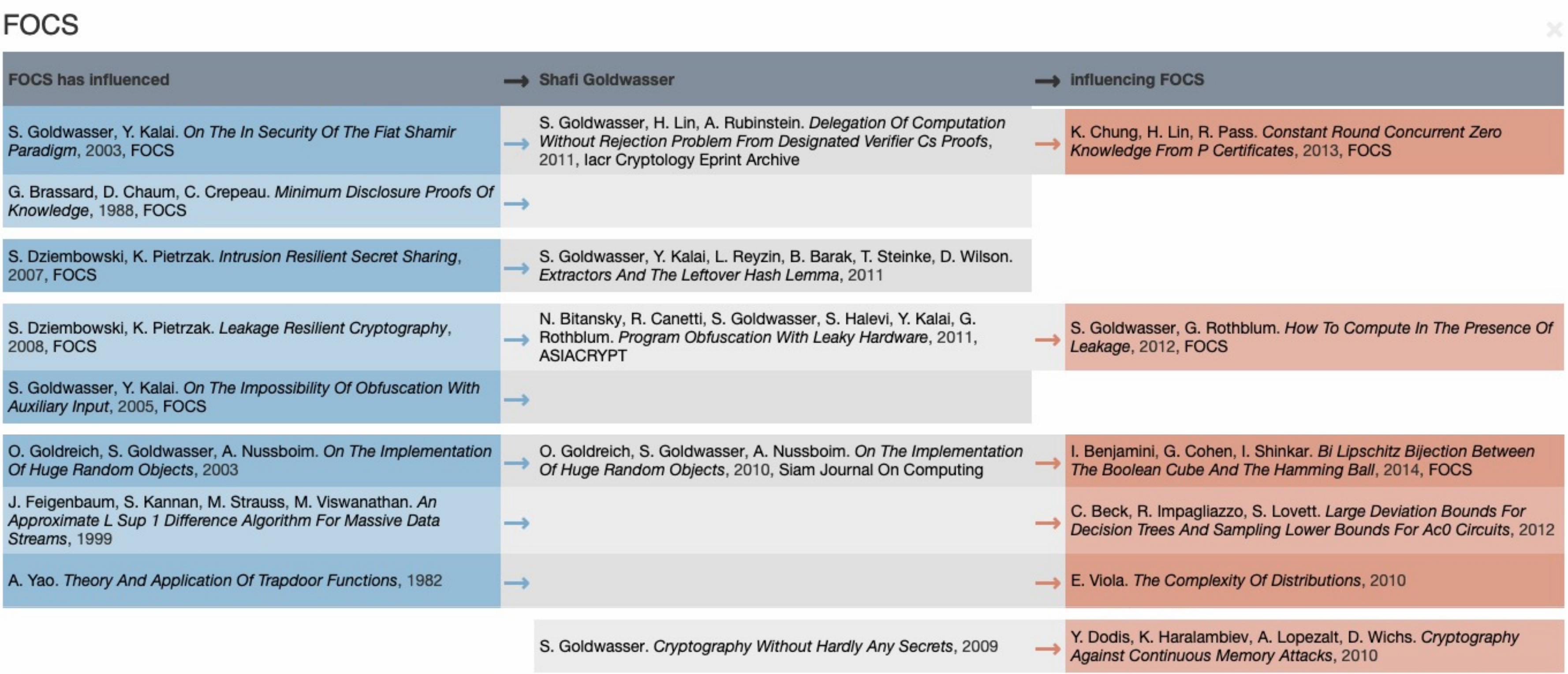}
    \caption{Detail page showing the underlying influence between FOCS (Symposium on Foundations of Computer Science) and Shafi Goldwasser, generated by clicking the FOCS node in the \inflower \autoref{fig:interface_flower}(c). The blue and red columns present FOCS papers that have influenced and have been influenced by Goldwasser. Arrows indicate the direction of influence. The entries are chronologically ordered by the ego's own papers (grey column, middle).}
    \label{fig:detail_page}
    \postfigmoveup
\end{figure}

To understand what constitutes the influence in the flower, the system can show the underlying paper information.
For example, \autoref{fig:detail_page} presents the details page when clicking the FOCS (IEEE Symposium on Foundations of Computer Science) in the author-to-venue flower
of Shafi Goldwasser \autoref{fig:interface_flower}(c). The left (blue) column lists the papers of the alter node (FOCS) that have influenced the ego, Goldwasser. The middle column shows the papers of the ego. Finally, the last column lists the papers in FOCS and that are influenced by papers written by the ego.
The blue and red arrows represent the flow of influence between the pairs of papers.
The blue (red) arrows in the table combine to correspond to the blue (red) edge between two entities in the flower.

%% file: sections/evaluation.tex
\secmoveup
\section{Case studies}
\label{sec:cases}
\sectxtmoveup
Having described the design of the \inflower metaphor and the system for searching and interacting with the flower, we present five case studies for using \infmap: (1) the career of a researcher; (2) an example of delayed impact; (3) the intellectual heritage of a lab; (4) the evolution of a research community, and (5) using \inflowers beyond the academic context.

\secmoveup
\subsection{\inflower and a career trajectory}
\label{ssec:career}
\sectxtmoveup

\begin{figure}[t]
    \centering
    \includegraphics[width=0.48\textwidth]{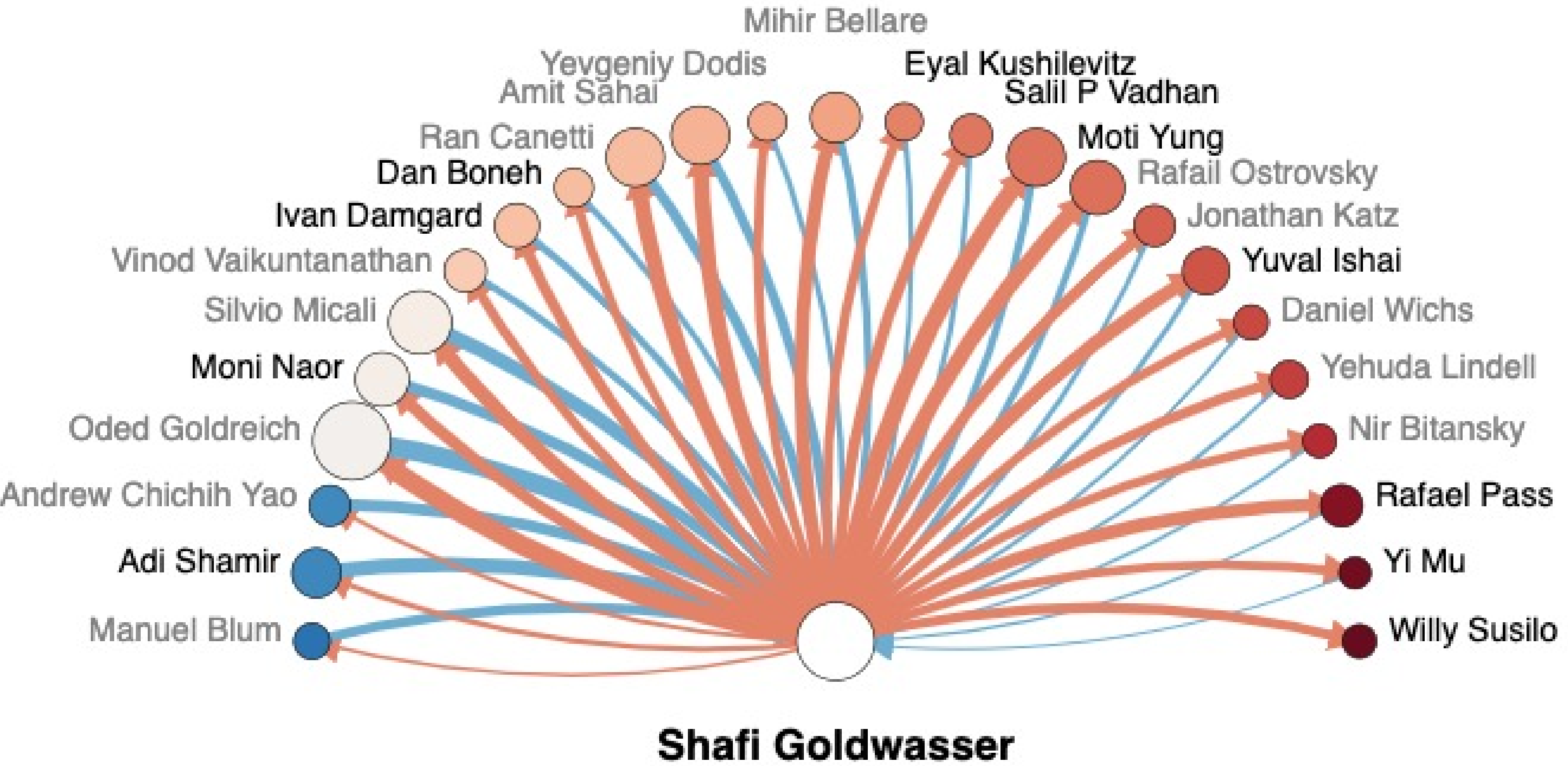}
    \caption{Author-to-author \inflower over Shafi Goldwasser's career 1982 -- 2017.
    See \autoref{ssec:career} for discussions.}
    \label{fig:shafi}\postfigmoveup
\end{figure}

We examine the \inflowers (\autoref{fig:interface_flower} and~\ref{fig:shafi}) of Shafi Goldwasser, who laid the foundations for modern cryptography (with Silvio Micali), and was recognised with a Turing Award in 2012. We compare the people and publication venues in her \inflower with those present in her publication profile and mentioned in her interview transcripts~\cite{shafi_interviews}, finding a few notable correspondences.  


We discuss three notable researchers in Goldwasser's author \inflower (\autoref{fig:shafi}). 
Manuel Blum was the PhD advisor for both Goldwasser and Micali, and his course on computational number theory was described by Goldwasser as ``a turning point''~\cite{shafi_interviews}. In the flower, Blum's influence on Goldwasser is larger than that of Goldwasser to Blum. 
Silvio Micali has been Goldwasser's long term collaborator from 
both being graduate students at UC Berkeley to faculty members at MIT. Between Micali and Goldwasser, the ratio of influences in the flower is close to one (off-white node color).
Amit Sahai is one of Goldwasser's PhD advisees, his node is a light shade of red, indicating that Goldwasser has had more influence on him than the other way round. 
In Goldwasser's author-to-venue flower \autoref{fig:interface_flower}(c), we can see that Goldwasser's work between 1996 and 2011 has influenced several communities that she does not publish in herself (dark venue names), such as information theory and communication networks.

Here, \infmap provides a detailed view of one's intellectual influences by different entity types, and the observations are corroborated by the biography 
and opinions of the ego.

\secmoveup
\subsection{A paper with delayed recognition}
\label{ssec:sleepingbeauty}
\sectxtmoveup

A paper with delayed recognition, often called a sleeping beauty or a Mendel syndrome, indicates a publication which received very little or no attention for a while, then was discovered later and received many citations. For example, the term Mendel syndrome is from Gregor Mendel in plant genetics, whose work is accepted by the community after three decades. 
There are many recognised causes for delayed recognition, such as conservatism of the community, errors in judgements, or lag time for the discipline to mature \cite{costas2011mendel}. Many works have investigated delayed recognition in the scientific literature \cite{costas2011mendel, van2004sleeping, ke2015defining}. 

\begin{figure}[t]
    \centering
    \includegraphics[width=0.48\textwidth]{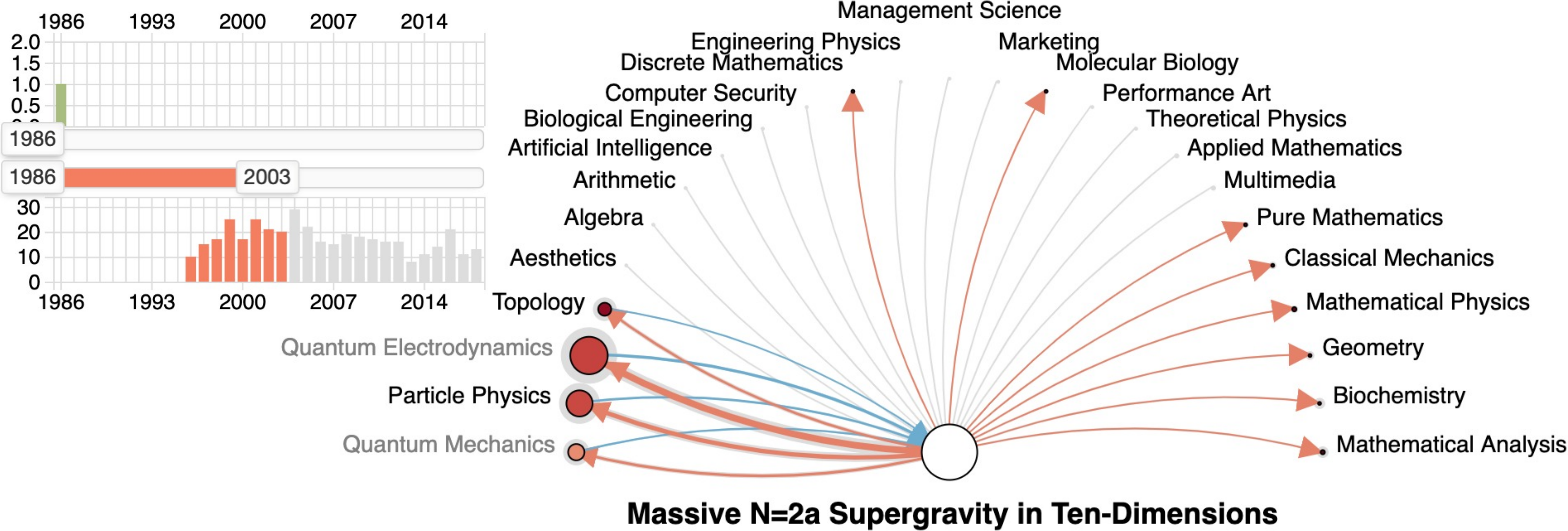}
    \caption{Paper-to-topic \inflower of a physics paper with delayed recognition \cite{romans1986massive} showing 1986 -- 2003 with the anchor flower 1986 -- 2018. The paper received the first citation after 10 years. Most of the initial citations are from physics, but later it gets more citations from different research fields.
    See \autoref{ssec:sleepingbeauty}.
    }
    \label{fig:sleeping}\postfigmoveup
\end{figure}

We examine a physics paper with delayed recognition\cite{romans1986massive}. When the term sleeping beauty was first introduced by \cite{van2004sleeping}, this paper was used as an extreme example. 
\autoref{fig:sleeping} presents the paper-to-topic flower of the paper, comparing the contrast flower from 1986 to 2003 with the anchor. The paper does not draw any attention until it gets its first citation after 10 years (1995). From 1995 to 2003, the paper influenced quantum electrodynamics and other physics and mechanics fields.
In 2004, the paper is introduced as an example of a sleeping beauty.
Since then, it is cited from entirely different fields identifying sleeping beauty papers such as management science, biological engineering, and marketing.
At the same time, it is still influential to physics.

\infmap provides the visualisation of delayed recognition by showing the delaying time and volume of attention, along with the changes in the influence profile.
This analysis  not only applies to papers with delayed recognition, but also to any paper, project (e.g. Appendix \autoref{fig:decapo_project}), or other type of academic entities.

\secmoveup
\subsection{Interdisciplinary work of a research institute}
\label{ssec:santafe}
\sectxtmoveup

\begin{figure}[tb]
    \centering
    \includegraphics[width=0.48\textwidth]{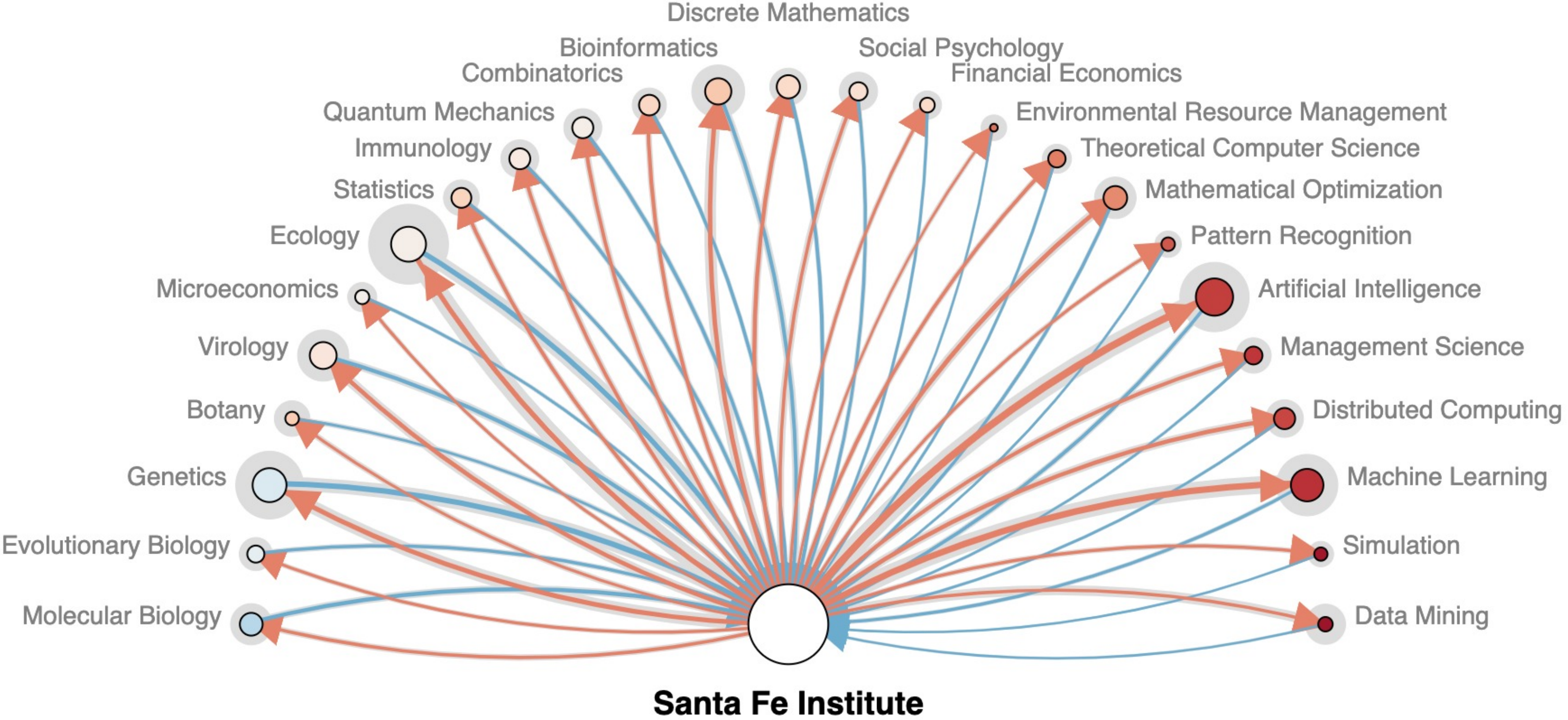}
    \caption{Institution-to-topic \inflower of the \href{https://en.wikipedia.org/wiki/Santa_Fe_Institute}{Santa Fe Institute} 1988 -- 2008, with the anchor flower 1988 -- 2018. 
    See \autoref{ssec:santafe}.
    }
    \label{fig:santa_fe}\postfigmoveup
\end{figure}



We examine the \href{https://en.wikipedia.org/wiki/Santa_Fe_Institute}{Santa Fe Institute}, an independent nonprofit research and education centre, in its first 20 years. The institute 
led the development of the field of complexity science -- 
the inquiry for
common mechanisms that lead to complexity in various theoretic and real world systems. 
In computer science, Santa Fe researchers are known for important results in network science. 
From the \inflower of the institute, we seek evidence not only 
on its multidisciplinary heritage, 
but also on the disciplines 
that it has influenced.

The study of complexity science examines many different domains, including computational, biological, and social systems. This is reflected in the \inflower (\autoref{fig:santa_fe}) by the inclusion and relatively equal size of the alter nodes: from botany, to social psychology, to simulations. 
In the \inflower, strong heritage of complexity science from biology is reflected as the two highest influencing alter nodes being ecology and genetics. Further, on the right hand side, the \inflower shows that the topic areas they have influenced are within the computer science side of complexity science -- artificial intelligence, machine learning, etc. 
From the anchor flower, the observation patterns of the influence is consistent in Santa Fe's current research, with interactions between biology and computing topics increasing the most over the past decade. 
The ratio and volume of total influence may also be affected by the increased publication volume on the corresponding topics in computing, ecology, and genetics.

In this case study, we use \infmap to profile 
the heritage and contributions a research centre has towards multiple fields.
We choose the Santa Fe Institute due to its moderate size and concentration of topics. In a large organisation (e.g. a university) the influence may be too spread out to see meaningful patterns. Instead, moderate-sized institutions can be established and grown in a decade or so, 
a timeframe relevant to decision-makers. 


\secmoveup
\subsection{Trend change of a venue}
\label{ssec:www}
\sectxtmoveup



\begin{figure}[t]
  \centering
  \includegraphics[width=0.46\textwidth]{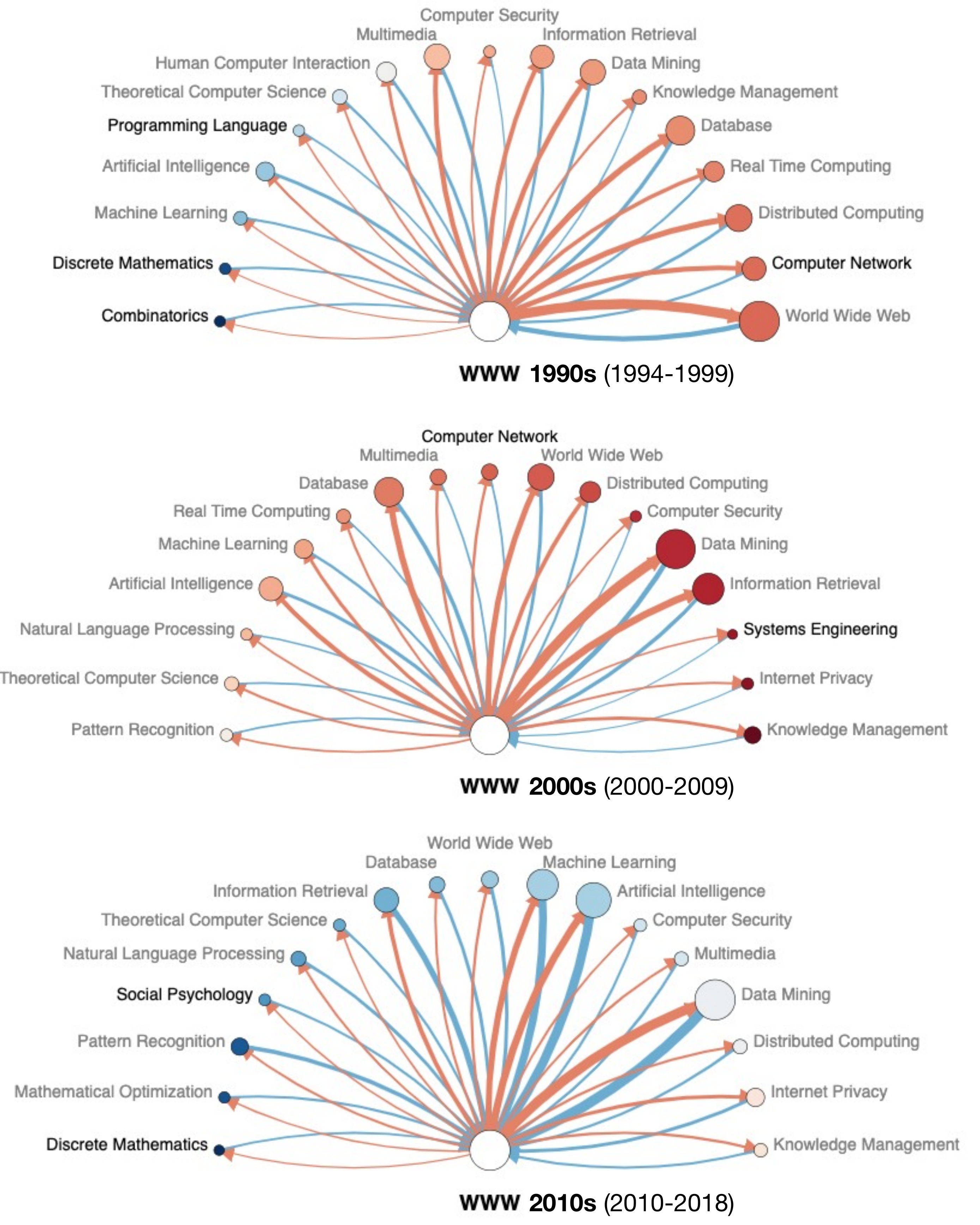}
  \caption{Venue-to-topic \inflowers of International World Wide Web Conference (WWW). Each flower represents the trends of the conference by selecting both the publications and citations of each decade -- 1990s, 2000s, and 2010s.
  See \autoref{ssec:www}.
  }
  \label{fig:www_trends}\postfigmoveup
\end{figure}

Computer science as a field has been having increasingly many publications every year and the fast development of ideas has made studying this evolution important \cite{hoonlor2013trends}.
We look at the International World Wide Web Conference (WWW) in this case study to consider how \infmap can be used to examine a conference which captures the changing landscape of computer science. In our analysis, we use a series of \inflowers to examine the change in topics over time in WWW (\autoref{fig:www_trends}). We divide the time span of WWW into three non-overlapping time intervals, 1994-1999 (1990s), 2000-2009 (2000s), and 2010-2018 (2010s). Each \inflower is created using publications and citations of the corresponding time interval.
 The anchor flower is not utilised in this example, as we are more concerned about the ratio of influence in each topic area, rather than the total growth overtime. The topics of papers, rather than the topic of influence, can be found in \autoref{fig:www_topic_breakdown} in the appendix.

\autoref{fig:www_trends} presents three venue-to-topic \inflowers of WWW in the 1990s, 2000s, and 2010s. Contrary to the namesake of the conference, the relative influence of WWW to the topic of the world wide web has been
decreasing. The world wide web topic has the highest incoming and outgoing influence in the 1990s, but has decreased in the subsequent years. Further, the colouring of the nodes shows that the influence direction between WWW and the world wide web topic has been changed. Contrasting to the red in the 1990s flower, the alter node in the 2010s flower has become blue indicating that WWW gets influenced more by the topic than its influence on the topic. With the decrease in the world wide web topic, there is also a notable increase in relative influence for data mining playing a major role from the 2000s. 
The machine learning and artificial intelligence topics have emerged over time, becoming particularly prominent in the 2010s flower. 
One explanation is that synergy between WWW and artificial intelligence and machine learning grew stronger. These observations are corroborated by the appendix \autoref{fig:www_topic_breakdown}, where we can see that strength of world wide web topic in published papers decreases, whilst data mining and machine learning increase. 

In this case study we have demonstrated how one can use \infmap to examine the changes in research topics of a publication venue. 
\infmap visualises the primary topics of influence in a venue.
We note that this analysis is not limited to conferences or journals, but to any group or community that publishes over time.

\secmoveup
\subsection{\inflowers of social media data}
\label{ssec:twitter}
\sectxtmoveup

\begin{figure}[t]
  \centering
  \includegraphics[width=0.45\textwidth]{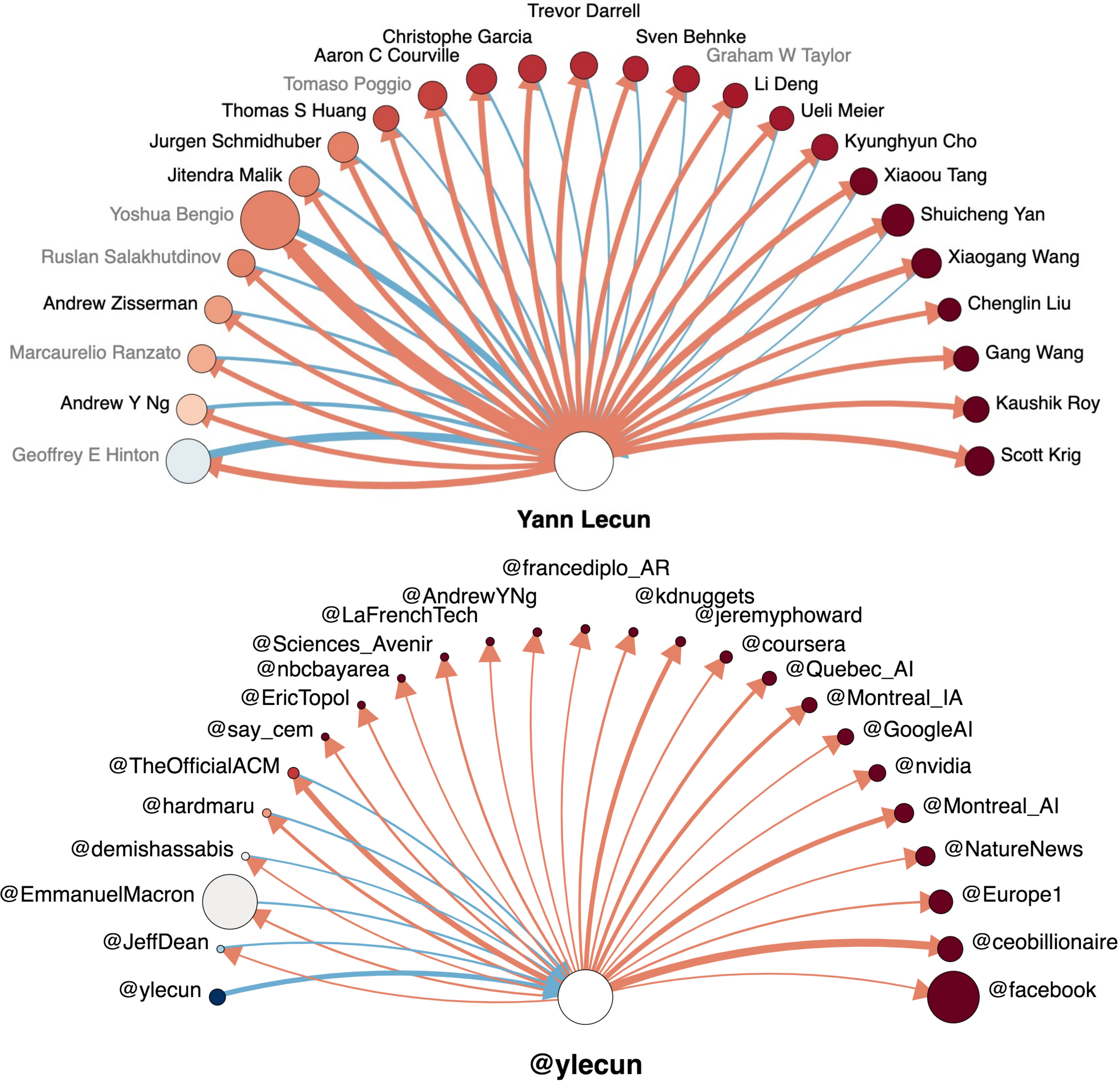}
  \caption{Academic (top) vs Twitter (bottom) \inflowers for Yann LeCun, 2018 Turing co-awardee for breakthroughs in deep neural networks. 
  See \autoref{ssec:twitter}.
  }
  \label{fig:twitter_flower}\postfigmoveup
\end{figure}

The \inflower metaphor can be applied to data outside of academic graphs.
\autoref{fig:twitter_flower} presents two \inflowers for Yann LeCun, one of the 2018 Turing Awardees who has an active Twitter profile. On the top is the author-to-author \inflower over LeCun's career, as defined in \autoref{sec:measuring} and similar to \autoref{fig:shafi}. The bottom flower is generated using one week of tweets mentioning LeCun (or Twitter user {\em @ylecun}, the ego node) until two days after the award announcement on 2019-03-27.
The interacting Twitter users form the alter nodes.
The node sizes are proportional to their respective number of Twitter followers. The edges describe the influence made by retweets, mentions, and replies. The thickness of edges are proportional to the number of times one user has mentioned the other.
Similarly to the academic flowers, the influence flow is opposite to the direction of mentions. The alters mentioning LeCun produces a red edge to show that LeCun influenced the alter.


LeCun's academic flower shows co-recipients Hinton and Bengio having the most influence, both are reciprocated. Moreover, Andrew Ng, who is well-known in machine learning research and active on Twitter, is the only common node across both flowers.
On the Twitter flower, three types of nodes mentioned {\em @ylecun}:
people in the artificial intelligence industry ({\em @demishassabis, @JeffDean, @jeremyhoward, @hardmaru)}, organisations in France and Canada - LeCun and Bengio's home countries ({\em @EmmanuelMacron, @Sciences\_Avinir, @LaFrenchTech, @Montreal\_AI}), and industry and scientific organisations ({\em @facebook, @GoogleAI, @nvidia, @NatureNews, @TheOfficialACM}).
The few reciprocating edges are to {\em @TheOfficialACM} who conferred the award, the French President Macron, and a few others in the AI industry.


Contrasting the academic and social media \inflowers, we can see that the two data sources capture activities in very different communities (research vs media) and the actors of influence
tend to be very different. More generally, this case study illustrates that \inflower can be readily adapted for other data domains and that one can
rethink the mapping between flower elements and data attributes, e.g. follower count becomes the additional attribute reflected in node size.




%% file: sections/conclusion.tex
\secmoveup
\section{Conclusion}
\label{sec:conclusion}
\sectxtmoveup

We propose the \inflower, a new visual metaphor 
to portray the flow of influence between academic entities. 
We develop the \infmap system to interactively search entities and create influence flowers, supported by efficient indexing of the entire Microsoft Academic Graph. 
Further, we present case studies illustrating the use of the \inflower in understanding 
one's scientific career, the dynamics of interdisciplinary impact, 
the intellectual profile of an institution, 
and the topic shift of an academic community. 
We also discuss an example \inflower using social media data. 

Future work lies in two fronts.
In this visual design, we would like to ease the tension between its visualisation function and richness of the flower metaphor, such as to endow meanings to size and shade of the petals, and to enact growth.
On the functional front, we would like more versatile comparisons of flowers beyond subsets in time, 
and using the \inflower to understand the production and consumption of science at large. 

%% file: sections/appendix.tex
\appendix
\section*{Appendix to VAST 2019 paper ``Influence Flowers of Academic Entities''}

\section{Indexing and improving scalability}
\label{ssec:indexing}


This project intends to build not only prototype visualisations, but also a system usable by anyone exploring intellectual influence. Since the MAG dataset contains millions of entities and billions of relations, efficient indexing is crucial for enabling interactive querying of different types of academic entities.
We use the Elasticsearch software stack for data storage, full-text search, and a web interface. We index both the raw data and generated cache entries. 
We use one virtual machine to host Elasticsearch server. It takes roughly two months to index and update the entire dataset. 


The time it takes to query and compute a new \inflower is roughly proportional to the total number of papers related to the queried entity (authored, referenced, and subsequently cited by others). 
The system needs to do relational queries on all cited and citing papers to obtain their authors, institutions, topics, publication years and venues. We now use an example to illustrate the difficulty of performing this on-the-fly.
The VAST conference (\autoref{fig:main_teaser}) has 678 papers in MAG with 8,238 references and 8,378 citations. 
One query is needed to obtain all papers by this conference ID. For each of the 678 publications, we run three queries to obtain reference and citation links, authors and affiliations, and topic information. Two queries are required for the up to 16,616 (citations possibly overlap) linked papers to get authors, affiliations and topics. In total, up to 35,267 queries are required to generate the VAST \inflower, taking 4 minutes and 17 seconds to create when searched for the first time.
To improve the responsiveness, we cache relational queries involving author, affiliation, and topics (dubbed as a {\em partial} cache). For papers that are associated with the centre entity, the reference and citation links are also cached (dubbed as a {\em complete} cache).
Using the cache, the VAST flower is re-created with 679 queries, taking only 5 seconds.
Cache entries are automatically created from user searches.
Curating the \infmap gallery (\autoref{sec:gallery})
with a set of authors (including Turing award winners), venues and projects yielded 9 million paper cache entries, of which 3\% are complete and 97\% are partial caches.


\section{Influence score normalisation}
\label{sec:normalisation}

Upon closer examination of \autoref{eq:infscore} in \autoref{ssec:infscore}, 
\begin{equation}
    \bS^{(k,l)} = \bA^{(k)} \bC \bA^{(l)T} \tag{1}
\end{equation}
we will find that if one citation is made in journal $j$ to paper $i$ with $K$ authors, the total amount of influence score generated is $K$. This presents an apparent contradiction with one citation being the unit of influence -- the amount of influence should not change arbitrarily due to the size of author teams (in modern scientific practice, this can be a solo author, or thousands of scientists in a large consortium).

We consider three mutually exclusive normalisation schemes:
\vspace{-\topsep}
\begin{itemize}
    \setlength{\parskip}{0pt}
    \setlength{\itemsep}{0pt plus 1pt}
    \item[(s1)] normalising the association matrix $\bA$ by the number of entities on the cited paper,
    \item[(s2)] normalising the citation matrix $\bC$ by the number of references in the citing paper,
    \item[(s3)] normalising the transpose of the association matrix $\bA^T$ by the number of entities on the citing paper.
\end{itemize}
\vspace{-\topsep}

Due to the lack of a ground truth value of influence to compare these definitions to, we investigate the eight combinations of these weightings empirically. 
We provide the author-to-author \inflowers of the two researchers in computer science, both of whom have more than 15 years of academic career with more than 2500 citations in MAG. They were asked which of the definitions produced flowers that 
agrees with their own impressions
of who they have influenced and been influenced by. 
Using (s2) is considered to be the least accurate. Authors of shorter papers (that tend to have fewer reference items) have 
inflated scores since each citation is inversely weighted by the number of references.
Also, the inverse weighting makes influence scores are difficult to interpret. 
Schemes (s2) and (s3) ascribed influence to authors who the researchers being influenced did not recognise; investigation revealed that the authors were minor contributors to highly influential work. Scheme (s1) addresses this, but as an author writes a paper with more co-authors, their measure of influence reduces. 
No normalisation is the second most favoured scheme. 
Among the other combinations, researchers 
agree the most
with the set of outer nodes when (s1) alone was used, i.e., normalising $A$ by the number of entities on the cited paper.

\begin{equation}
    \bar{A}_{ij} = { n(A_{ij}) \over {\sum_{k} n(A_{kj})} },~~ \bar{A}_{ij} \; \in [0,1]^{|E| \times |P|}
    \label{eq:normalisation}
\end{equation}
\autoref{eq:normalisation} describes the general form of normalising $A$ by the proportion of entities on the cited paper, where $n(A_{ij})$ is the number of appearances of entity $e_i$ in paper $p_j$.
The intuition for this normalisation 
varies slightly based on entity type. For author and topic entities, influence is equally divided by the total number of entities. For institution entity, influence is proportional to the number of entity appearances in a paper (e.g., 3 of 5 authors from MIT). For venue type, no normalisation is needed because a paper can be published to at most one venue \footnote{Note that 0.2\% of papers (426,506) have both publishing journal and conference information. We do not split the influence in this case.}. We use the normalised influence matrix $\bar{\bS}^{(k,l)} = \bar{\bA}^{(k)}\bC\bA^{(l)T}$ to create the \inflower, where $\bar{\bS} \in \mathbb{R}^{|E^{(k)}| \times |E^{(l)}|}$.

In addition to the empirical study, we demonstrate to twelve other senior researchers in different fields their normalised \inflowers. 
Most researchers 
recognise a large number of names in the flower, 
especially all nodes with incoming influence (whom they cited), 
and bigger nodes with outgoing influence (who cited them significantly).
We also noticed variations due to differing citation practices in different fields. 

Various methods for normalising citation metrics exist \cite{ioannidis2016citation}, including journal-level normalisation based on ranking or impact factor. However, 
using prestige of a venue as a proxy for the quality 
of a paper has raised concerns~\cite{berger2019goto, hicks2015bibliometrics}.
The \inflower does not replace any citation metrics or propose a single index to compare, but instead provides a tool with which to understand influence.

\section{Browsing lists of pre-curated entities}
\label{sec:gallery}

\begin{figure*}[h]
  \centering
  \includegraphics[width=0.65\textwidth]{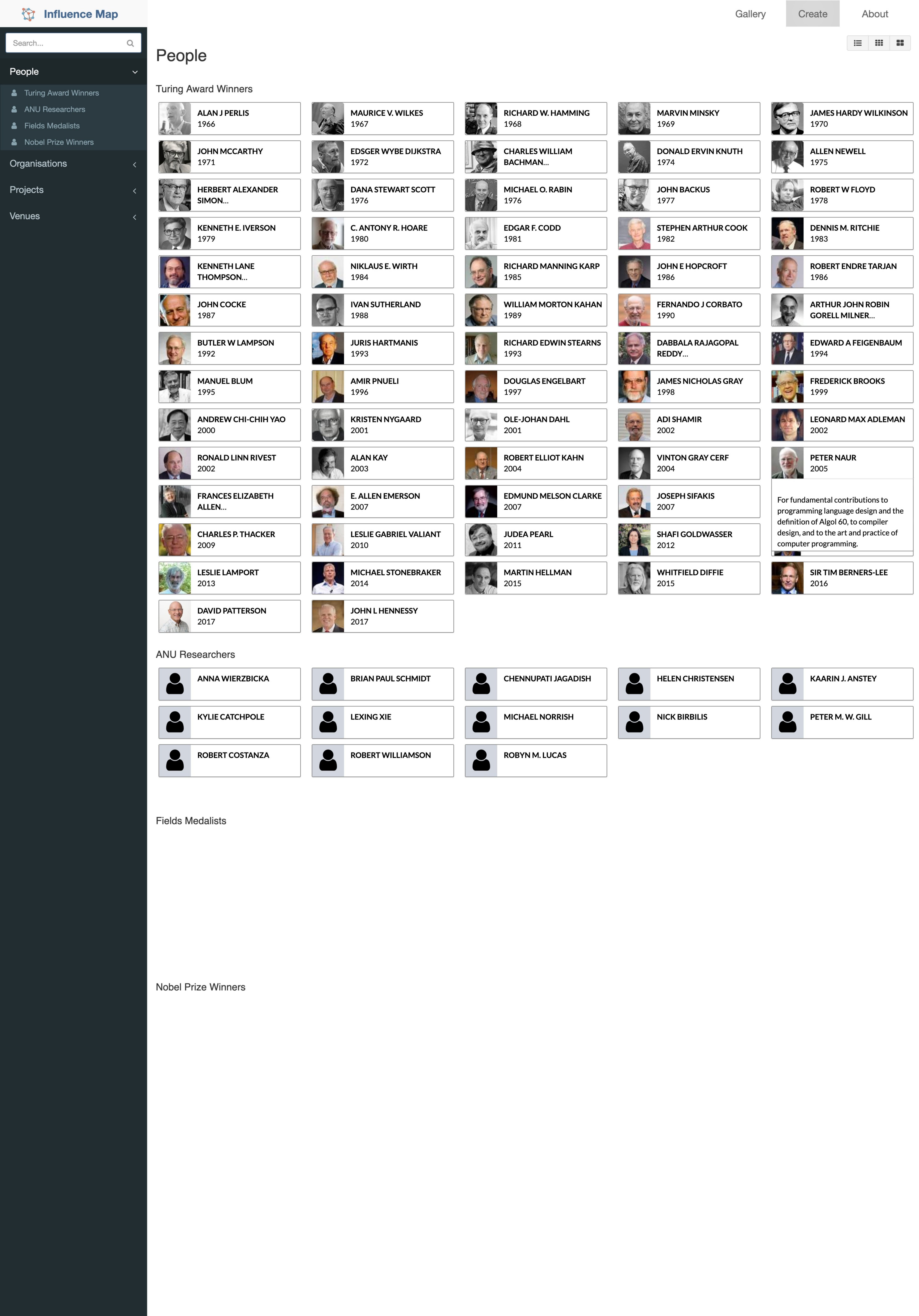}
  \caption{Snapshot of the gallery page containing Turing award winners. The gallery includes different types of precurated academic entities, such as people, organisations, projects, and venues. Clicking an entity opens the influence flower of the entity.}
  \label{fig:inf_gallery}
\end{figure*}

\infmap provides two entry options; browsing pre-curated \inflowers and creating a new \inflower by searching. 
Creating an \inflower requires users prior knowledge of the query entity. Browsing the gallery gives newcomers idea of the overall dataset and the types of academic entities supported by the system. \autoref{fig:inf_gallery} shows the snapshot of the \infmap gallery. Categories of the \inflowers are given on the left-hand side. Each of the categories contains pre-curated entities with a breif description. Clicking on an entity leads to the \inflower page of the entity.

\section{Computing relevance score for academic entities}
\label{ssec:relevance}
We use Elasticsearch for searching names of academic entities, and sort the search result by their relevance score. Elasticsearch computes a relevance score using a similarity model based on term frequency and inverse document frequency. The higher the score, the more relevant to the query. 
We observe
that the score is between 10 and 25 for the \infmap system.
We modify the scoring function to favour entities with higher citation counts: $score = (score)^3 \times log(N_c + \epsilon)$, where $N_c$ is the number of citations and $\epsilon$ is constant. The log of the citation count is used to ensure that large entities with slightly different names are not missed. The addition of $\epsilon$ is to account for relevant entities with 0 or few citations.
The $score$ is raised to a higher power to increase its significance.

\section{Additional figures}

\begin{figure*}[h]
  \centering
  \includegraphics[width=0.55\textwidth]{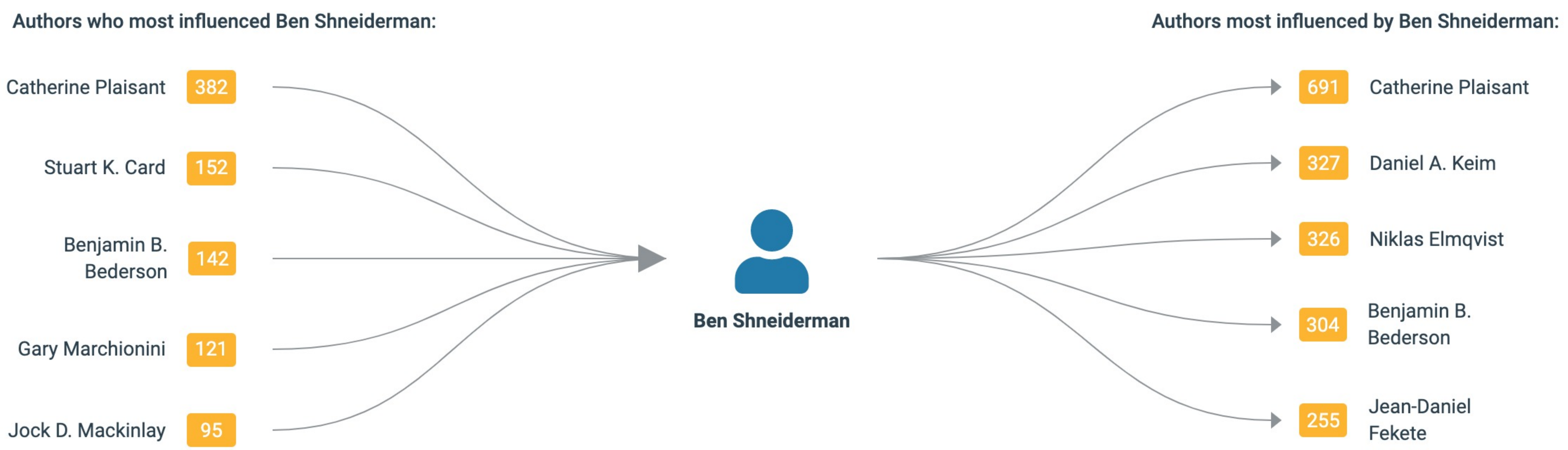}
  \caption{Semantic Scholar has an author influence pane to show the top 5 other authors with the highest incoming and outgoing influence, respectively. The influence score is calculated based on a weighted combination of citations and Highly Influential Citations\cite{valenzuela2015identifying}. Snapshot retrieved July 2019.}
  \label{fig:semantic_scholar}
\end{figure*}

\begin{figure*}[h]
  \centering
  \includegraphics[width=0.55\textwidth]{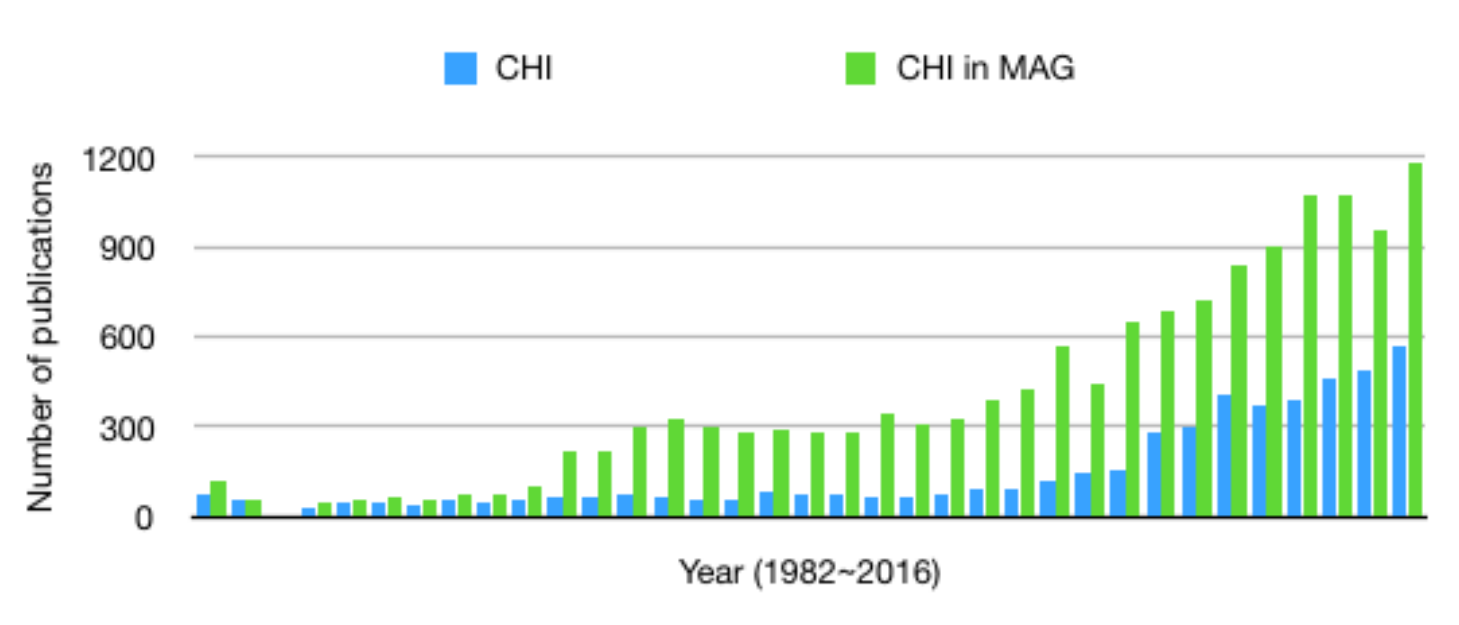}
  \caption{Statistic of ACM SIGCHI papers from 1982 to 2016, as seen by statistics on github (https://github.com/steveharoz/Vis-Acceptance-Rates/blob/master/acceptance rates.csv, retrieved April 2019) and in MAG. The number of actual proceedings and the number of papers indexed by MAG differ by up to 3 times due to the large numbers of short ancillary articles including demos and posters.}
  \label{fig:chi_papercount}
\end{figure*}

\begin{figure*}[h]
  \centering
  \includegraphics[width=0.55\textwidth]{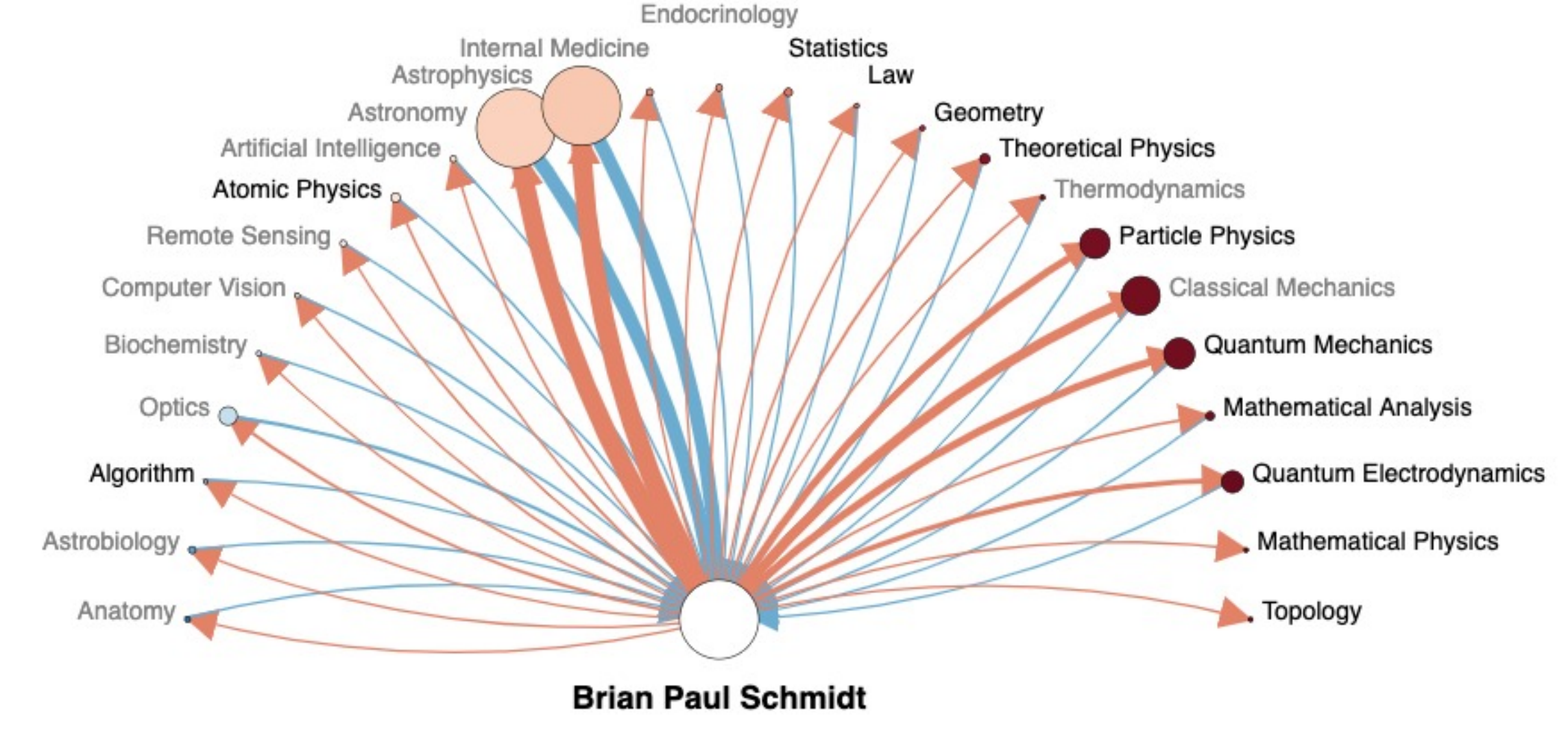}
  \vspace{1mm}
  \includegraphics[width=0.6\textwidth]{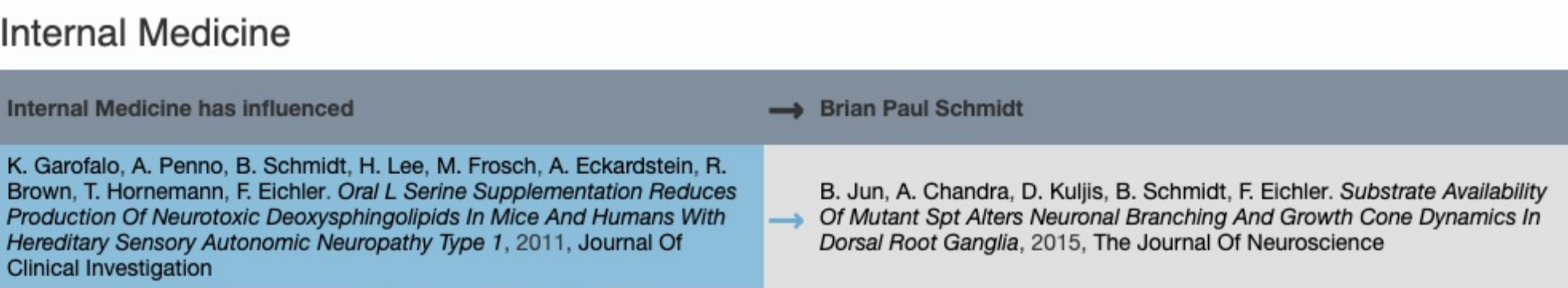}
  \caption{The author-to-topic \inflower of Brian P. Schmidt, an astronomer at the Research School of Astronomy and Astrophysics at the Australian National University. The \inflower presents that his main area of influence are astronomy and astrophysics, but also shows that Schmidt has influenced and been influenced by internal medicine. This is caused by an error in entity resolution with another Brian P. Schmidt, a neuroscientist in Massachusetts General Hospital.}
  \label{fig:brian_medicine}
\end{figure*}

\begin{figure*}[h]
  \centering
  \includegraphics[width=0.55\textwidth]{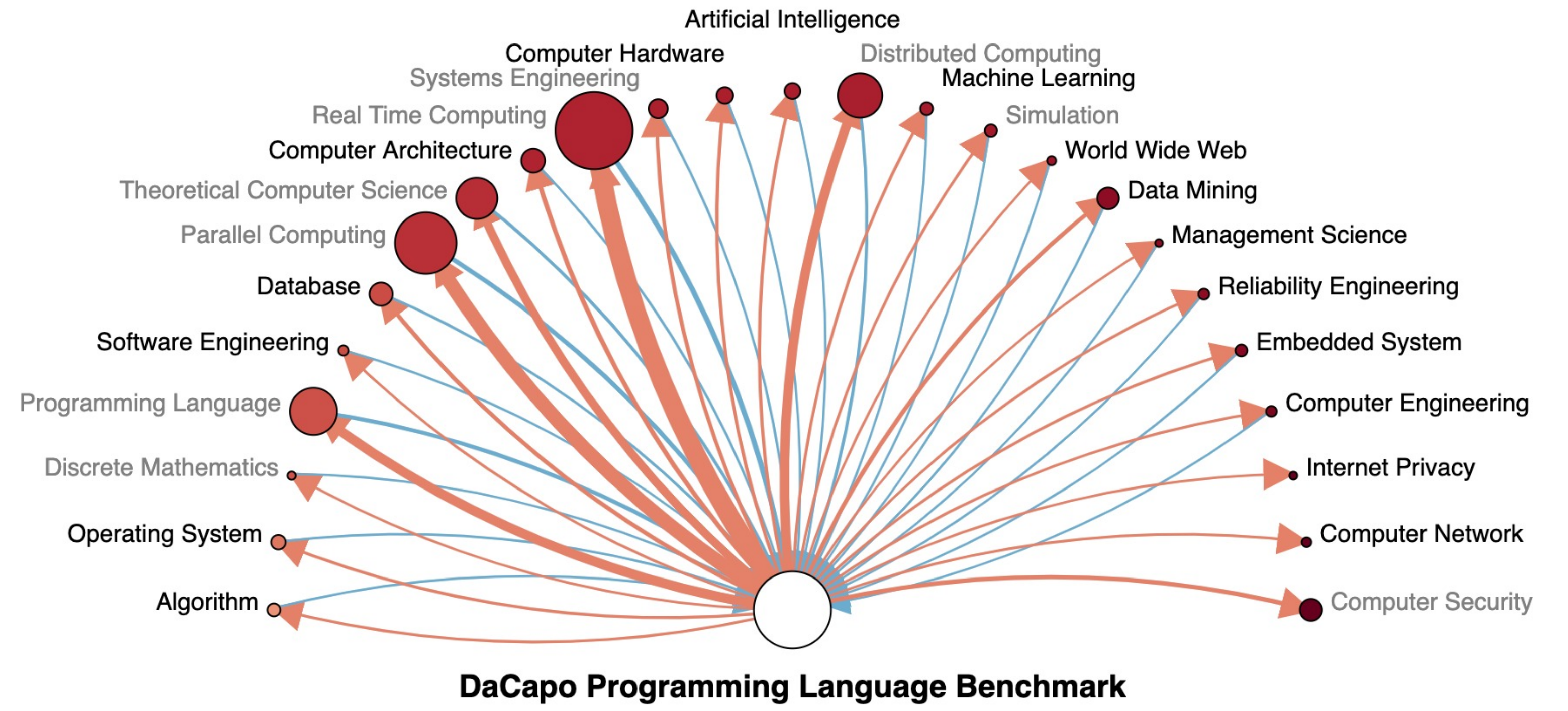}
  \caption{The project-to-topic \inflower of the \href{http://dacapobench.org/}{DaCapo benchmark project}. The DaCapo benchmark suite is a tool for Java benchmarking. It is developed by the programming language, memory management, and computer architecture communities over eight years at eight institutions. This \inflower is generated from 16 research papers written as part of the project from 2001 to 2009. The DaCapo project has highly influenced the research area where the project started, such as programming language, parallel computing, real-time computing, and distributed computing (topics in grey colour). The project has also expanded its influence to other fields (in black colour), such as database, artificial intelligence, and computer network.} 
  \label{fig:decapo_project}
\end{figure*}

\begin{figure*}[h]
  \centering
  \includegraphics[width=0.6\textwidth]{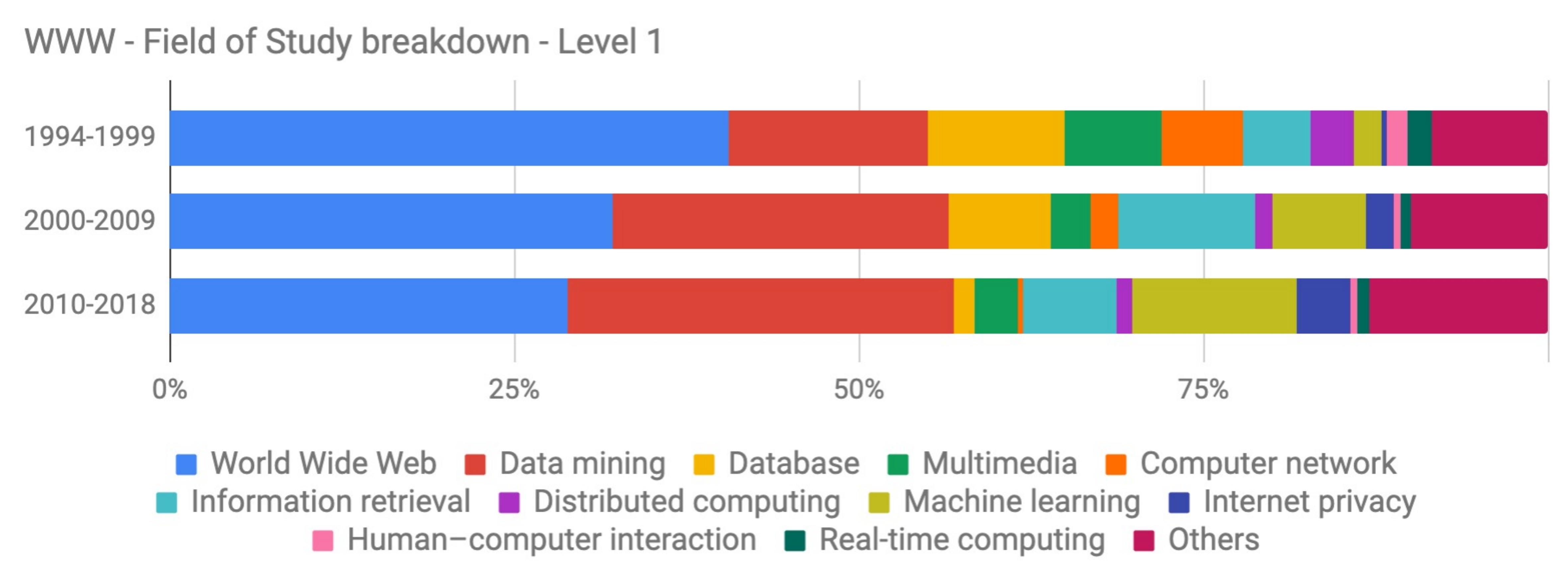}
  \caption{Level 1 topic breakdown of the International World Wide Web Conference (WWW). This bar chart shows the percentage of paper topics published in WWW from 1994 to 2018, broken down into three non-overlapping time periods. The percentage of the world wide web topic has been constantly decreasing over time and data mining topic is gradually increasing. We observe the growth of machine learning and internet privacy while database and multimedia shrink.
  The similar trends in incoming and outgoing influence can be seen in \autoref{fig:www_trends}.}
  \label{fig:www_topic_breakdown}
\end{figure*}

This section includes five additional figures that complement discussions in the main text. \autoref{fig:semantic_scholar} shows a snapshot of the author influence pane
in the Semantic Scholar, which consists of the top 5 influencers and influencees of an author (\autoref{sec:related}). \autoref{fig:chi_papercount} shows the difference in the number of ACM SIGCHI proceedings and the number in MAG (\autoref{sec:dataset}). \autoref{fig:brian_medicine} presents an example \inflower due to the imperfect entity resolution in MAG (\autoref{sec:dataset}). \autoref{fig:decapo_project} shows that the \inflower can be used to visualise the influence of a project (\autoref{ssec:sleepingbeauty}). \autoref{fig:www_topic_breakdown} presents the topics of papers published in WWW change over time (\autoref{ssec:www}).

